\documentclass[11pt,a4paper]{article}
\usepackage[top=2.2cm, bottom=2.2cm, left=2.2cm, right=2.2cm]{geometry}
\usepackage{graphicx,color}
\usepackage{amsmath,amssymb,amscd}
\usepackage{caption}

\newtheorem{theorem}{Theorem}
\newtheorem{theorema}{Theorem}
\newtheorem{theoremb}{Theorem}

\newtheorem{cor}[theorema]{Corollary}

\newtheorem{prop}[theoremb]{Proposition}
\newenvironment{proof}[1][Proof]{\textbf{#1.} }{\qed \vspace{5pt}}

\newcommand\qed{\phantom{\underline{y}}\hfill\hfill$\square$}
\newcommand{\comm}[1]{}

\newcommand\C{{\mathbb C}}

\newcommand\g{\mathfrak{g}}

\newcommand\op[1]{\mathop{\rm #1}\nolimits}

\newcommand\p{\partial}

\newcommand\po{$\!\!\!{\text{\bf.}}$ }

\newcommand\R{{\mathbb R}}

\newcommand\bx{{\boldsymbol x}}

\begin{document}

\vspace{-2cm}
\title{Second-order PDEs in 3D with Einstein-Weyl conformal structure}

\author{S. Berjawi$^1$, E.V. Ferapontov$^{1,2}$,
B.S. Kruglikov$^{3,4}$, V.S. Novikov$^{1,2}$}
     \date{}
     \maketitle
\begin{center}
$^1$Department of Mathematical Sciences, Loughborough University \\
Loughborough, Leicestershire LE11 3TU, United Kingdom \\
    \ \\
$^2$Institute of Mathematics, Ufa Federal Research Centre\\
Russian Academy of Sciences, 112 Chernyshevsky Street \\
Ufa 450008, Russian Federation\\
    \ \\
$^3$Department of Mathematics and Statistics,
Faculty of Natural Sciences\\
UiT the Arctic University of Norway,
Troms\o\ 90-37 Norway\\
    \ \\
$^4$Department of Mathematics and Natural Sciences \\
University of Stavanger, 40-36 Stavanger, Norway\\
    [2ex]
e-mails: \\[1ex]
\texttt{S.Berjawi@lboro.ac.uk}\\
\texttt{E.V.Ferapontov@lboro.ac.uk}\\
\texttt{Boris.Kruglikov@uit.no} \\
\texttt{V.Novikov@lboro.ac.uk}\\

\end{center}

\medskip

\begin{abstract}

Einstein-Weyl geometry is  a triple  $(\mathbb{D},g,\omega)$ where $\mathbb{D}$ is a symmetric connection, $[g]$ is a  conformal structure and $\omega$ is a covector such that
 \begin{itemize}
\item connection $\mathbb {D}$ preserves the conformal class $[g]$,
that is, $\mathbb{D}g=\omega g$;
\item trace-free part of the symmetrised Ricci tensor of
$\mathbb {D}$ vanishes.
\end{itemize}

Three-dimensional Einstein-Weyl  structures naturally arise on solutions of  second-order dispersionless integrable PDEs in 3D. In this context, $[g]$  coincides with the characteristic conformal structure and is therefore uniquely determined by the equation. On the contrary, covector $\omega$ is a somewhat more mysterious object, recovered from the Einstein-Weyl conditions.

We demonstrate that, for generic second-order PDEs (for instance,
for all equations not of Monge-Amp\`ere type), the covector $\omega$ is also expressible in terms of the equation, thus providing an efficient `dispersionless integrability test'.  The knowledge of $g$ and $\omega$ provides a dispersionless Lax pair by an explicit  formula which is apparently new.

Some partial classification results of  PDEs with Einstein-Weyl characteristic conformal structure are  obtained. A rigidity conjecture is proposed according to which for any generic second-order PDE with Einstein-Weyl property, all dependence on the 1-jet variables can be eliminated via a suitable contact transformation.
\medskip

\medskip

\end{abstract}

\newpage

\section{Introduction}

We consider second-order partial differential equations (PDEs) in 3D,
\begin{equation}
F(x^i, u, u_i, u_{ij})=0,
\label{F}
\end{equation}
where $u$ is a scalar function  of the three independent variables
$x^0,x^1,x^2$, and we denote $u_i=u_{x^i}$, $u_{ij}=u_{x^ix^j}$, etc.
For every solution of equation \eqref{F} the corresponding
characteristic variety,
 $$
\sum_{i\leq j}\frac{\partial F}{\partial u_{ij}}p_ip_j=0,
 $$
defines a conformal structure $g=g_{ij}dx^idx^j$ where the symmetric matrix  $\bigl(g_{ij}\bigr)_{3\times3}$ is inverse to the matrix of the symbol
$\bigl(g^{ij}\bigr)_{3\times3}$, with
$g^{ij}=\frac{1+\delta_{ij}}2\frac{\partial F}{\partial u_{ij}}$
(no summation).
Here and in what follows we assume the nondegeneracy condition
$\det g^{ij}\not\equiv0$, i.e.\ $[g]$ is well-defined on a generic solution of \eqref{F}. Equations with nondegenerate characteristic
variety will be called nondegenerate.

We will be interested in equations (\ref{F}) whose characteristic conformal structure $g$   satisfies the Einstein-Weyl property on every solution of (\ref{F}) (PDEs with EW property for short). Recall that  Einstein-Weyl geometry is defined by a triple
$(\mathbb{D}, g, \omega)$ where $\mathbb{D}$ is a symmetric connection, $g$ is a conformal structure and $\omega$ is a covector such that  \cite{Cartan}:
 \begin{itemize}
\item[(a)] connection $\mathbb {D}$ preserves the conformal class $[g]$: $\mathbb{D}g=\omega g$;
\item[(b)] trace-free part of the symmetrised Ricci tensor of $\mathbb {D}$ vanishes.
 \end{itemize}
In coordinates, this gives
 \begin{equation}
\mathbb{D}_kg_{ij}=\omega_k g_{ij}, ~~~ R_{(ij)}=\Lambda g_{ij},
\label{EW}
 \end{equation}
where $\omega= \omega_kdx^k$ is a covector, $\mathbb{D}_k$ denotes covariant derivative, $R_{(ij)}$ is the symmetrised Ricci tensor of $\mathbb{D}$,  and $\Lambda$ is some function.  In fact one needs to specify $g$ and $\omega$ only, then the first set of equations (\ref{EW}) uniquely defines the corresponding Weyl connection $\mathbb{D}$.
We will  refer to $\omega$ as the Weyl covector. It was shown in \cite{FKr} that, for  broad classes of  translationally invariant equations (\ref{F}),  the Weyl covector  is expressed in terms of  $g$ by the  explicit formula
 \begin{equation}
\omega_k=2g_{kj}\mathcal{D}_{x^s}(g^{js})+\mathcal{D}_{x^k}(\ln\det g_{ij}),
\label{omega}
 \end{equation}
where $\mathcal{D}_{x^k}$ denotes the total derivative with respect to $x^k$:
 $$
\mathcal{D}_{x^k}=\partial_{x^k}+u_k\partial_u+u_{ik}\partial_{u_i}+
u_{ijk}\partial_{u_{ij}}+\dots.
 $$
We emphasize  that in the general  case  formula (\ref{omega}) is no longer valid. Finding a universally applicable  formula for the Weyl covector is one of the main objectives of this paper.
Since the characteristic conformal structure $g$ depends on the 2-jet of $u$, one can show that the Weyl covector  depends on no more than 3-jets,
and is linear in the third-order partial derivatives of $u$.
(Recall that the $k$-jet of $u$ at a point $x$ can be identified
with the collection of partial derivatives $\p^\nu u$ of order $|\nu|\leq k$.)

\medskip

\noindent{\bf Remark.} Given a three-dimensional conformal background $[g]$, the problem of reconstruction of a covector $\omega$ such that $(g,\omega)$ satisfiy the Einstein-Weyl equations is far from trivial: it was shown in \cite{ETn, ET3} that, for given $g$, Einstein-Weyl equations reduce to a complicated differential system for $\omega$. This system is overdetermined, and not in involution. Thus, it
may be inconsistent (the cases of general left-invariant metric on $S^3$,  metric `Sol', or any `sufficiently generic' $g$), or possess multiple solutions (the case of flat $g$, and some other metrics with multiple
Killing vectors).
In particular, there is no explicit `formula' for $\omega$ in terms of $g$.
What makes difference in our case is that we are dealing with a whole family of conformal structures parametrised by solutions $u$ of  second-order PDE (\ref{F}). The corresponding Einstein-Weyl equations split in the higher-order derivatives of $u$, thus providing additional constraints for $\omega$, both  differential and algebraic. This leads to a formula for $\omega$ depending on the equation, and involving 3-jets of  $u$ only,
with the leading part given by (\ref{omega}).

\medskip

We recall that Einstein-Weyl equations  (\ref{EW}) are  integrable
by twistor-theoretic methods \cite{Hitchin}; in \cite{DFK}
this was explicitly demonstrated in the Manakov-Santini gauge.
PDEs (\ref{F}) satisfying EW property can be viewed as reductions of the Einstein-Weyl equations. This, in particular, implies the existence of a dispersionless Lax representation  \cite{CalKrug}.
We recall that  dispersionless Lax pair consists of two parameter-dependent vector fields $\hat X, \hat Y$ for which the integrability condition
$$
[\hat X, \hat Y]\in {\rm span} \langle \hat X, \hat Y\rangle
$$
holds identically modulo  (\ref{F}).
Relations of dispersionless integrable systems to
Einstein-Weyl geometry have been discussed in \cite{Ward, Calderbank,   Dun4, Dun6, Dun7, Dun3}.

\subsection{Summary of main results}

\noindent  {\bf Partial classification results.}  In Section \ref{sec:ex} we demonstrate that  EW property is an efficient classification/integrability criterion. To illustrate the approach we obtain  complete lists of PDEs with EW property within the following three classes:

\begin{itemize}

\item Dispersionless lattice equations
$$
u_{xy}=f(x, y, t, u, u_x, u_y, u_t, u_{tt}).
$$
 Modulo natural equivalence transformations preserving this class  there exists a unique example with $f_{u_{tt}, u_{tt}}\ne 0$, the so-called Boyer-Finley equation $u_{xy}=e^{u_{tt}}$ \cite{BF}. This example shows that  EW property is a rather stringent constraint.

\item Nonlinear wave equations
$$
u_{tt}=f(x, y, t, u, u_x, u_y, u_t)\, u_{xy}.
$$
The EW property leads to a generic case $f=\frac{\sinh^2u_t}{u_xu_y}$, plus a number of degenerations.

\item Generalised Dunajski-Tod equations
$$
(u_{tt}-u)u_{xy}-(u_{xt}-u_{x})(u_{yt}+u_{y})=f(x, y, t, u, u_x, u_y, u_t).
$$
The EW property leads to a generic case $f=c^2\frac{u_xu_y}{\cosh^2ct}$,
plus a number of degenerations.

\end{itemize}

\noindent For all equations arising in the classification we calculate the corresponding Einstein-Weyl structures and dispersionless Lax pairs. The structure of contact symmetry algebras indicates that all resulting equations are  contact non-equivalent.

\bigskip

\noindent {\bf Reconstruction of the Weyl covector.} In Section \ref{sec:omega} we outline a general procedure to calculate the covector $\omega$. This procedure applies to all equations that are {\it not of Monge-Amp\`ere type}, and gives an expression for $\omega$ in terms of the equation (see Section \ref{sec:MA} for  Monge-Amp\`ere conditions in 3D). We look for $\omega$ in the form
$$
\omega_k=\Omega_k+\phi_k
$$
where $\Omega_k$ are given by  formula (\ref{omega}), and $\phi_k$ are the `correction' terms. Substituting $g,\, \omega$ into  Einstein-Weyl conditions (\ref{EW}) and splitting the resulting equations in the third-order derivatives of $u$, we conclude that the correction terms $\phi_k$ must be functions of the 2-jet of $u$ only.  Furthermore, along with a number of differential relations, $\phi_k$ must satisfy an {\it algebraic} system of 20 linear inhomogeneous equations which, in the non-Monge-Amp\`ere case, determines $\phi_k$ uniquely.
In other words, the system has the form $A\phi=B$ where $A$
is a $20\times3$ matrix of $\op{rank}=3$,
and $B$ is a 20-component vector (both depend on the 2-jet of $u$).
We supply a Mathematica program which calculates the linear system, and the Weyl covector $\omega$. Summarising, we have the following result.

 \begin{theorem}\po \label{t1}
For every nondegenerate non-Monge-Amp\`ere equation (\ref{F}) with
EW property, the Weyl covector $\omega$ is algebraically determined
by the equation.
 \end{theorem}

\medskip

\noindent{\bf Remark.}
For Monge-Amp\`ere equations the matrix of the linear system $A$ and the vector $B$
vanish identically, and further analysis is required to reconstruct $\phi_k$. In fact, the EW conditions provide an overdetermined
differential system for $\phi$ which, in generic case, implies a formula for
$\phi$ through differential closure (compatibility analysis) of the system.
We demonstrate this with examples in Section \ref{sec:ex}.

\medskip

As a bi-product of our analysis we obtain a remarkable fact that, for any second-order PDE (\ref{F}) with EW property, `freezing' the 1-jet of $u$ (that is, giving the variables $x^i, u, u_i$ arbitrary constant values), results in an integrable Hirota type equation $F(u_{ij})=0$.

\bigskip

\noindent  {\bf General formula for  dispersionless Lax pair.} For equation (\ref{F}) with EW property, in Section \ref{sec:Lax} we propose an algorithm to calculate the corresponding dispersionless Lax pair. Here is a brief summary. Let $g$ and $\omega$ be the  characteristic conformal structure and the Weyl covector, respectively. Let us introduce the so-called null coframe $\theta^0, \theta^1, \theta^2$ such that
$$
g=4\theta^0\theta^2-(\theta^1)^2.
 $$
Let  $V_0, V_1, V_2$ be the dual   frame, and
let $c_{ij}^k$ be the structure functions defined by commutator expansions $[V_i,V_j]=c_{ij}^kV_k$.
The Lax pair is given by vector fields
 $$
\hat{X}=V_0+\lambda V_1+m\p_\lambda,\ \
\hat{Y}=V_1+\lambda V_2+n\p_\lambda,
 $$
where
 \begin{align*}
m=&
(\tfrac12c_{12}^1-\tfrac14\omega_2)\lambda^3
+(\tfrac12c_{02}^1-c_{12}^2-\tfrac12\omega_1)\lambda^2
+(\tfrac12c_{01}^1-c_{02}^2-\tfrac14\omega_0)\lambda-c_{01}^2,\\[8pt]
n=&
-c_{12}^0\lambda^3
+(\tfrac12c_{12}^1-c_{02}^0+\tfrac14\omega_2)\lambda^2
+(\tfrac12c_{02}^1-c_{01}^0+\tfrac12\omega_1)\lambda
+(\tfrac12c_{01}^1+\tfrac14\omega_0);
 \end{align*}
here $\omega_i$ are the components of the Weyl covector: $\omega=\omega_i\theta^i$. In combination with Theorem \ref{t1} we have the following result.

 \begin{theorem}\po \label{t2}
Every nondegenerate second-order PDE with EW property is integrable,
and the dispersionless Lax pair is algebraically determined by
the Weyl covector $\omega$ and the function $F$ of (\ref{F}).
 \end{theorem}

\begin{cor} \label{cor}
For every nondegenerate non-Monge-Amp\`ere equation (\ref{F}) with EW property, the dispersionless Lax pair is algebraically determined by the equation.
\end{cor}

This result sounds, in a sense, surprising: intuition coming from the theory of soliton equations tells us that  reconstruction of a Lax pair for a given PDE (known to be integrable) should require `integration' of some kind.

\bigskip

\noindent {\bf Rigidity conjecture.} In Section \ref{sec:rig} we formulate a   rigidity conjecture which states that, in the  non-Monge-Amp\`ere case, every  PDE (\ref{F}) with EW property can be reduced to a dispersionless Hirota form $F(u_{ij})=0$ via a suitable contact transformation. In other words, all dependence on the 1-jet variables $x^i,u,u_i$ can be eliminated  (for Monge-Amp\`ere equations this is not true).

To illustrate this phenomenon we consider a PDE \cite{MaksEgor}
 $$
u_{tt}=\frac{u_{xy}}{u_{xt}}+\frac{1}{6}\varphi(u_{xx})u_{xt}^{2},
 $$
for which  EW property is equivalent to the  Chazy equation $\varphi'''+2\varphi \varphi''-3\varphi'^2=0$.
We prove that any deformation of the form
 $$
u_{tt}=\frac{u_{xy}}{u_{xt}}+\frac{1}{6}f(x, u, u_x, u_{xx})u_{xt}^{2},
 $$
which satisfies EW property, is trivial (contact-equivalent to the undeformed equation). We believe that our method of proof can be extended to the general case.

 \section{Examples and classification results}
 \label{sec:ex}

Given a class of second-order PDEs in 3D,  we impose  Einstein-Weyl conditions for the characteristic conformal structure $g$  to  obtain  classification results. This procedure can be viewed as a `dispersionless integrability test', and  is manifestly contact-invariant.
Some illustrative examples are given below. We emphasize that in all examples the Weyl covector $\omega$, as well as the associated dispersionless Lax pair,  are expressible  in terms of the  equation by explicit formulae that work for all special cases arising in the classification.

\subsection{Dispersionless lattice equations }
\label{sec:lat}

Here we consider equations of the form
\begin{equation}\label{lat}
u_{xy}=f(x, y, t, u, u_x, u_y, u_t, u_{tt}).
\end{equation}
In the translationally invariant case, such equations arise  as dispersionless limits of integrable lattices
$$
u^n_{xy}=F(u, u_x, u_y, u^{n-1}, u^n, u^{n+1}),
$$
see \cite{FHKN}. The  characteristic conformal structure of equation (\ref{lat}) has the form
$$
g=4f_{u_{tt}}dxdy-dt^2.
$$
Assuming $f_{u_{tt},u_{tt}}\ne 0$ (which is equivalent to the requirement that equation (\ref{lat}) does not belong to the Monge-Amp\`ere class), one can show that  the Weyl covector is given by the following formula in terms of the right-hand side $f$:
$$
\omega=\left(\frac{2}{3}\frac{f_{u_t}}{f_{u_{tt}}} +\frac{10}{3}\frac{\mathcal{D}_t(f_{u_{tt}})}{f_{u_{tt}}}
-\frac{4}{3}\frac{\mathcal{D}_t(f_{u_{tt},u_{tt}})}{f_{u_{tt},u_{tt}}} \right)dt
$$
where $\mathcal{D}_t$ denotes the total $t$-derivative.
The requirement that $g,\, \omega$ satisfy Einstein-Weyl conditions on every solution of equation (\ref{lat}) leads to a system of differential constraints (integrability conditions) for the right-hand side $f$, the simplest of them being
 $$
\begin{array}{c}
f_{u_{tt},u_{tt},u_{x}}=\frac{f_{u_{tt},u_{tt}}f_{u_{tt},u_{x}}}{f_{u_{tt}}}, \quad f_{u_{tt},u_{tt},u_{y}}=\frac{f_{u_{tt},u_{tt}}f_{u_{tt},u_{y}}}{f_{u_{tt}}}, \quad f_{u_{tt},u_{tt},u_{t}}=\frac{f_{u_{tt},u_{tt}}f_{u_{tt},u_{t}}}{f_{u_{tt}}},
 \\[1em]
f_{u_{tt},u_{x},u_{x}}=\frac{f_{u_{tt},u_{tt}}f_{u_{x},u_{x}}}{f_{u_{tt}}}, \quad f_{u_{tt},u_{x},u_{y}}=\frac{f_{u_{tt},u_{x}}f_{u_{tt},u_{y}}}{f_{u_{tt}}}, \quad f_{u_{tt},u_{y},u_{y}}=\frac{f_{u_{tt},u_{tt}}f_{u_{y},u_{y}}}{f_{u_{tt}}},
 \\[1em]
f_{u_{tt},u_{tt},u_{tt}}=\frac{f_{u_{tt},u_{tt}}^2}{f_{u_{tt}}},
\quad f_{u_{tt},u_{x},u_{t}}=\frac{f_{u_{tt},u_{x}}f_{u_{tt},u_t}}{f_{u_{tt}}},\quad
\quad f_{u_{tt},u_{y},u_{t}}=\frac{f_{u_{tt},u_{y}}f_{u_{tt},u_{t}}}{f_{u_{tt}}},
 \\[1em]
f_{u_{tt},u_{t},u_{t}}=\frac{f_{u_{tt},u_{t}}^2}{f_{u_{tt}}}+2(f_{u_{tt},u_x}f_{u_{tt},u_y}-f_{u_x,u_y}f_{u_{tt},u_{tt}}),
\end{array}
 $$
plus a number of more complicated constraints. Note that the set of integrability conditions is not in involution, and the prolongation implies further second-order relations such as
 $$
f_{u_{tt},u_{tt}}f_{u_t}=f_{u_{tt},u_t}f_{u_{tt}}, \quad
f_{u_{tt}, u_x}=f_{u_{tt}, u_y}=f_{u_{x}, u_x}=f_{u_{y}, u_y}=f_{u_{x},u_t}=f_{u_{y},u_t}=0,\quad \text{etc.}
 $$
Consequently, modulo the equivalence transformations $u\to U(u, x, y, t)$, equation \eqref{lat} has the form
 \begin{equation*}
u_{xy}= e^{u_{tt}+\varphi u_t+\frac{2}{9}u(3\varphi'+\varphi^2)}
 \end{equation*}
where $\varphi(t)$ is an arbitrary function. It can be set equal to zero
via a suitable transformation $t\to a(t),\ u\to b(t)u+c(t)$,
thus leading to the unique canonical form
 \begin{equation}\label{BFeq}
u_{xy}=e^{u_{tt}},
 \end{equation}
known as the Boyer-Finley (BF) equation \cite{BF}. This example demonstrates rigidity of the Einstein-Weyl requirement.

Integrable equations of type (\ref{lat})  possess a Lax representation $[\hat X, \hat Y]\in {\rm span} \langle \hat X, \hat Y\rangle$ with
 $$
\begin{array}{l}
\hat X=\partial_y+\lambda f_{u_{tt}} \partial_t +\lambda^2 \left(\frac{2}{3}f_{u_{tt}}\frac{\mathcal{D}_t f_{u_{tt}, u_{tt}}}{f_{u_{tt}, u_{tt}}}-\frac{5}{3}\mathcal{D}_tf_{u_{tt}}-\frac{1}{3}f_{u_t} \right) \partial_{\lambda}, \\[1em]
\hat Y=\lambda \partial_x+\partial_t+\lambda \left(\frac{1}{3}\frac{f_{u_t}}{f_{u_{tt}}}+\frac{2}{3}\frac{\mathcal{D}_t f_{u_{tt}}}{f_{u_{tt}}}-\frac{2}{3}\frac{\mathcal{D}_t f_{u_{tt}, u_{tt}}}{f_{u_{tt}, u_{tt}}}-\lambda \frac{\mathcal{D}_x  f_{u_{tt}}}{f_{u_{tt}}}  \right) \partial_{\lambda}.
\end{array}
 $$
Remarkably,  this Lax pair works modulo integrability conditions satisfied by $f$ and is therefore fully invariant under the equivalence transformations preserving  class (\ref{lat}). For BF equation \eqref{BFeq} it simplifies to
 $$
\begin{array}{c}
\hat X=\partial_y+\lambda e^{u_{tt}} \partial_t -\lambda^2 e^{u_{tt}}u_{ttt} \partial_{\lambda}, \quad
\hat Y=\lambda \partial_x+\partial_t-\lambda^2 u_{ttx} \partial_{\lambda}.
\end{array}
$$

\subsection{Nonlinear wave equations}
\label{sec:wave}
Here we consider quasilinear equations of the form
\begin{equation}
u_{tt}=f(x, y, t, u, u_x, u_y, u_t)\,u_{xy}.
\label{nwe}
\end{equation}
The characteristic conformal structure is
$$
g=\frac{4}{f}dxdy-dt^2,
$$
the corresponding Weyl covector is given by
$$
\omega=(-2{\mathcal{D}}_t \ln f+\varphi(t))\, dt.
$$
At this stage, $\varphi(t)$ is some function to be determined.
The requirement that $g,\,\omega$ satisfy Einstein-Weyl conditions on every solution of equation (\ref{nwe}) leads to a system of differential constraints (integrability conditions) for the right-hand side $f$. The simplest of them are as follows:
 $$
\begin{array}{c}
f_{u_x,u_x}=2\frac{f_{u_x}^2}{f}, \quad f_{u_x,u_y}=\frac{f_{u_x}f_{u_y}}{f}, \quad f_{u_y,u_y}=2\frac{f_{u_y}^2}{f},\\[1em]
f_{u_x,u_t}=\frac{f_{u_x}f_{u_t}}{f}, \quad f_{u_y,u_t}=\frac{f_{u_y}f_{u_t}}{f}, \quad f_{u_t,u_t}=\frac{ff_{u_t}^2-2f_{u_x}f_{u_y}}{f^2}, \\[1em]
f_{u_x,u}=\frac{f_{u_x}f_{y}+u_yf_uf_{u_x}-ff_{y, u_x}}{u_yf}, \quad
f_{u_y,u}=\frac{f_{u_y}f_{x}+u_xf_uf_{u_y}-ff_{x, u_y}}{u_xf},
\end{array}
 $$
plus  four more complicated constraints that  involve $\varphi(t)$.
One of them is
 $$
\begin{array}{c}
\frac{3}{2}f_{u_t} (\varphi f-2f_t-2u_tf_u)+\frac{u_x}{u_y}(ff_{y, u_x}-f_yf_{u_x})+ \frac{u_y}{u_x}(ff_{x, u_y}-f_xf_{u_y})
+2ff_{t, u_t}\\[1em]
+2u_tff_{u, u_t}+2f_xf_{u_x}+u_xf_uf_{u_x}-ff_{x, u_x}+2f_yf_{u_y}+u_yf_uf_{u_y}-ff_{y, u_y}=0.
\end{array}
 $$
Analysis of these constraints shows that for (nonlinear) integrable equations  \eqref{nwe} the coefficient $f_{u_t}$ cannot equal zero, and
we obtain an explicit formula for $\varphi(t)$ in terms of $f$:
 $$
\begin{array}{l}
\varphi(t)= 2\frac{f_t}{f}+2u_t\frac{f_u}{f}-\frac{2}{3ff_{u_t}}\frac{u_x}{u_y}(ff_{y,u_x}
-f_yf_{u_x})-\frac{2}{3ff_{u_t}}\frac{u_y}{u_x}(ff_{x, u_y}-f_xf_{u_y}) \\[1em]
-\frac{2}{3ff_{u_t}}(2ff_{t, u_t}+2u_tff_{u, u_t}+2f_xf_{u_x}+u_xf_uf_{u_x}-ff_{x, u_x}+2f_yf_{u_y}+u_yf_uf_{u_y}-ff_{y, u_y}).
\end{array}
 $$
It is a non-trivial corollary of the integrability conditions that the right-hand side of this expression is a function of $t$ only. In any case, we have an explicit  formula for $\omega$ in terms of the equation.

Solving the integrability conditions results in the following generic case:
 $$
u_{tt}=\frac{\sinh^2u_t}{u_xu_y}\, u_{xy},
 $$
as well as a number of singular strata.
Normal forms are achieved modulo equivalence transformations $x\to \eta (x), \ y\to \psi(y), \ u\to u+tp(x)+tq(y)+r(x)+s(y)$,  translation of
the $t$-variable, rescaling of $u$, discrete symmetries
$x\leftrightarrow y$ and $t\mapsto-t$, and the transformation
$(x,y,t,u)\mapsto(x,y,1/t,u/t)$, which all leave the class  (\ref{nwe}) form-invariant.
The final list is summarised below.

 \begin{center}
\begin{tabular}{|l|l|l|}
 \hline
\# & Equation \eqref{nwe} & Contact symmetry algebra $\g$\\
 \hline
1\vphantom{$\frac{\frac{A}A}{\frac{A}A}$} &
$f=\frac{\sinh^2u_t}{u_xu_y}$ & % parametrized by
2 functions of 1 variable \& 3 constants\\
 \hline
2\vphantom{$\frac{\frac{A}A}{\frac{A}A}$} &
$f=\frac{u_t^2}{u_xu_y}$ &  % parametrized by
3 functions of 1 variable \& 2 constants\\
\hline
3\vphantom{$\frac{\frac{A}A}{\frac{A}A}$} &
$f=u_y^{-1}e^{u_t}$ &  % parametrized by
3 functions of 1 variable  \& 2 constants\\
\hline
4\vphantom{$\frac{\frac{A}A}{\frac{A}A}$} &
$f=u_y^{-1}$ &  % parametrized by
4 functions of 1 variable \& 2 constants\\
\hline
5\vphantom{$\frac{\frac{A}A}{\frac{A}A}$} &
$f=t^{-3/2}e^{u_t}$ &  % parametrized by
4 functions of 1 variable  \& 1 constant\\
\hline
6\vphantom{$\frac{\frac{A}A}{\frac{A}A}$} &
$f=e^{u_t}$ &  % parametrized by
4 functions of 1 variable  \& 2 constants\\
\hline
\end{tabular}
\captionof{table}{Complete list of (nonlinear) integrable cases.}
 \end{center}

Cases 1-6 are contact non-equivalent: while some of the symmetry algebras have the same
dimensional characteristics,
none are isomorphic as follows from the Lie algebra structure.
In all cases $\g$ is the right extension of an infinite ideal
by a Lie algebra of dimension $\leq 3$:
 $$
0\to\mathfrak{s}_\infty\longrightarrow\g\longrightarrow
\mathfrak{s}_\diamond\to0.
 $$
Below we describe $\g$ via $\mathfrak{s}_\infty=\langle Z_i\rangle$,
$\mathfrak{s}_\diamond=\langle V_j\rangle$ for each item of the table.
It turns out that in all cases, $\mathfrak{s}_\infty$ is the derived algebra
$[[\g,\g],[\g,\g]]$, and $\mathfrak{s}_\diamond$ is a subalgebra-complement. Thus, both the functional dimension and the number of
constants are invariantly defined.

\medskip

(1) The generators are $Z_1(a)=a(x)\p_x$, $Z_2(b)=b(y)\p_y$,
$V_1=t\p_t+u\p_u$, $V_2=\p_t$, $V_3=\p_u$.
Thus, $\mathfrak{s}_\infty=\op{Vect}(\R)\oplus\op{Vect}(\R)$,
$\mathfrak{s}_\diamond=\R\ltimes\R^2$ and
$[\mathfrak{s}_\infty,\mathfrak{s}_\diamond]=0$.

\medskip

(2) Here $Z_1(a)=a(x)\p_x$, $Z_2(b)=b(y)\p_y$,
$Z_3(c)=c(u)\p_u$, $V_1=t\p_t$, $V_2=\p_t$.
Thus, $\mathfrak{s}_\infty=\op{Vect}(\R)\oplus\op{Vect}(\R)\oplus\op{Vect}(\R)$,
$\mathfrak{s}_\diamond=\R\ltimes\R$ and
$[\mathfrak{s}_\infty,\mathfrak{s}_\diamond]=0$.

\medskip

(3) Here $Z_1(a)=a(x)\p_x+a'(x)t\p_u$, $Z_2(b)=b(x)\p_u$, $Z_3(c)=c(y)\p_y$,
$V_1=x\p_x+t\p_t+u\p_u$, $V_2=\p_t$. Thus,
$\mathfrak{s}_\infty=\mathfrak{s}_\infty'\oplus\mathfrak{s}_\infty''$ with
$\mathfrak{s}_\infty'=\op{Vect}(\R)\ltimes C^\infty(\R)$ and $\mathfrak{s}_\infty''=\op{Vect}(\R)$,
$\mathfrak{s}_\diamond=\R\ltimes\R$ and
in addition $[\mathfrak{s}_\infty,\mathfrak{s}_\diamond]
=\mathfrak{s}_\infty'$.

\medskip

(4) Here $Z_1(a)=a(x)\p_x-(a'(x)u+\frac12t^2a''(x))\p_u$, $Z_2(b)=b(x)\p_u$, $Z_3(c)=c(x)t\p_u$,
$Z_4(d)=d(y)\p_y$, $V_1=t\p_t+2u\p_u$, $V_2=\p_t$. Thus,
$\mathfrak{s}_\infty=\mathfrak{s}_\infty'\oplus\mathfrak{s}_\infty''$ with
$\mathfrak{s}_\infty'=\op{Vect}(\R)\ltimes(C^\infty(\R)\oplus C^\infty(\R))$ and $\mathfrak{s}_\infty''=\op{Vect}(\R)$,
$\mathfrak{s}_\diamond=\R\ltimes\R$ and
in addition $[\mathfrak{s}_\infty,\mathfrak{s}_\diamond]
=C^\infty(\R)\oplus C^\infty(\R)\subset\mathfrak{s}_\infty'$.

\medskip

(5) Here $Z_1(a)=a(x)\p_x+a'(x)t\p_u$, $Z_2(b)=b(x)\p_u$,
$Z_3(c)=c(y)\p_y+c'(y)t\p_u$, $Z_4(d)=d(y)\p_u$,
$V_1=t\p_t+(u-\frac12t)\p_u$. Thus,
$\mathfrak{s}_\infty=\mathfrak{s}_\infty'\oplus\mathfrak{s}_\infty''/
\langle Z_2(1)=Z_4(1)\rangle$ with
$\mathfrak{s}_\infty'=\op{Vect}(\R)\ltimes C^\infty(\R)=\mathfrak{s}_\infty''$,
$\mathfrak{s}_\diamond=\R$ and $[\mathfrak{s}_\infty,\mathfrak{s}_\diamond]
=C^\infty(\R)'\oplus C^\infty(\R)''\subset\mathfrak{s}_\infty$.

\medskip

(6) This has the same $\mathfrak{s}_\infty$ as in case (5),
but $\mathfrak{s}_\diamond=\R\ltimes\R$ is generated by
$V_1=2y\p_y+t\p_t+u\p_u$, $V_2=\p_t$.
In addition $[\mathfrak{s}_\infty,\mathfrak{s}_\diamond]
=C^\infty(\R)'\oplus \mathfrak{s}_\infty''$.

\bigskip

Equations from Table 1 possess a Lax representation $[\hat X, \hat Y]\in {\rm span} \langle \hat X, \hat Y\rangle$ with
 $$
\hat X=f\partial_y+\lambda \partial_t+\lambda^2(\mathcal{D}_t\log f-\tfrac12\varphi)\, \partial_{\lambda}, \quad
\hat Y=\lambda \partial_x+\partial_t+\lambda(\lambda\mathcal{D}_x\log f+\tfrac12\varphi)\, \partial_{\lambda},
 $$
here $\varphi(t)$ is the same as in the formula for the Weyl covector. Note that this Lax pair  works for all cases from Table 1 (upon substitution of the corresponding expression for $f$). In fact, one can say more: this Lax pair works modulo the integrability conditions satisfied by $f$, that is, it is invariant under the equivalence transformations used to obtain cases 1-6.

\medskip

\noindent{\bf Remark.}
Contact symmetry algebra of the BF equation, $u_{xy}=e^{u_{tt}}$,
contains 6 functions of 1 variable, therefore, it is not equivalent to any
of the items in Table 1.
Indeed, for the BF equation  we have:
$Z_1(a)=a(x)\p_x+\frac12a'(x)t^2\p_u$, $Z_2(b)=b(x)t\p_u$, $Z_3(c)=c(x)\p_u$,    $Z_4(d)=d(y)\p_y+\frac12d'(y)t^2\p_u$, $Z_5(e)=e(y)t\p_u$, $Z_6(f)=f(y)\p_u$,    $V_1=t\p_t+2u\p_u$, $V_2=\p_t$.
Therefore, we have
$\mathfrak{s}_\infty=\mathfrak{s}_\infty'\oplus\mathfrak{s}_\infty''$ where
$\mathfrak{s}_\infty'=\op{Vect}(\R)\ltimes (C^\infty(\R)\oplus C^\infty(\R))
=\mathfrak{s}_\infty''$;
$\mathfrak{s}_\diamond=\op{sol}(2)$ and
$[\mathfrak{s}_\infty,\mathfrak{s}_\diamond]=
(C^\infty(\R)\oplus C^\infty(\R))'\oplus
(C^\infty(\R)\oplus C^\infty(\R))''$.
Note though that BF equation can be obtained by potentiation from
case 6 of Table 1.

\subsection{Generalised Dunajski-Tod equations}
\label{sec:DT}

Here we consider Monge-Amp\`ere equations of the form
\begin{equation}
(u_{tt}-u)u_{xy}-(u_{xt}-u_{x})(u_{yt}+u_{y})=f(x, y, t, u, u_x, u_y, u_t).
\label{DT}
\end{equation}
For $f=4e^{2\rho t}$ this equation was discussed by Dunajski and Tod in the context of hyper-K\"ahler metrics with conformal symmetry \cite{Dun6}.
The characteristic conformal structure of equation (\ref{DT}) has the form
$$
g=(udt+u_xdx-u_{y}dy-du_{t})^2+4fdxdy.
$$
One can show that  the Weyl covector can be expressed in terms of the right-hand side $f$:
 \begin{equation}\label{EWforDT}
\omega =
2\Bigl(\frac{u_{xt}-u_x}{u_{tt}-u}dx-\frac{u_{yt}+u_{y}}{u_{tt}-u}dy\Bigr)+
2R\Bigl(dt+\frac{u_{xt}-u_x}{u_{tt}-u}dx+\frac{u_{yt}+u_{y}}{u_{tt}-u}dy\Bigr),
 \end{equation}
where $R=\frac{\mathcal{D}_tf}f$.
For $f=4e^{2\rho t}$  we have $R=2\rho$, which results in the Einstein-Weyl structure from \cite{Dun6}. The requirement that $g,\, \omega$ satisfy Einstein-Weyl conditions on every solution of equation (\ref{DT}) leads to a system of differential constraints (integrability conditions) for the right-hand side $f$. The simplest of them are as follows:
 $$
f_{u_x,u_x}=0, \quad f_{u_x,u_y}f-f_{u_x}f_{u_y}=0, \quad f_{u_y,u_y}=0,
 $$
plus a number of more complicated constraints. Solving the integrability conditions results in the  generic case
 $$
(u_{tt}-u)u_{xy}-(u_{xt}-u_{x})(u_{yt}+u_{y})=c^2\frac{u_xu_y}{\cosh^2ct}
 $$
(where $c$ is an arbitrary constant), as well as a number of other strata. Normal forms are obtained
modulo the following equivalence transformations:
$(x,y,t,u)\mapsto(\eta(x), \psi(y), t, u+p(x)e^t+q(y)e^{-t})$,
rescaling of $u$, translations of the $t$-variable, and
discrete symmetry $(x,y,t,u)\mapsto(y, x, -t, u)$, which all leave
the class  (\ref{DT}) form-invariant. Thus we obtain the following integrable cases:
 $$
 \begin{array}{c}
f=c^2\frac{(u_x+u_t+u)(u_y+u_t-u)}{\cosh^2c(x+y-t)}, \quad
f= c^2\frac{(u_x+u_t+u)u_y}{\cosh^2c(x-t)}, \quad
f= c^2\frac{u_xu_y}{\cosh^2ct},\quad
f= \frac{(u_t+u)(u_t-u)}{(x-y)^2}, \\[1em]
f= e^{ct}(xu_x-\frac{u_t+u}{c+1}), \quad
f= e^{-t}(u_x+u_t+u), \quad
f= e^{ct}u_x, \quad
f= e^t(u_t-u), \quad
f= e^{ct}.
 \end{array}
 $$

This list can be reduced further via point transformations as follows.
 \begin{itemize}
\item[] $(x,y,t,u)\mapsto(\frac12e^{2x},-\frac12e^{-2y},t-x-y,u\,e^{x-y})$
maps $f=c^2\frac{(u_x+u_t+u)(u_y+u_t-u)}{\cosh^2c(x+y-t)}$ to
$f=c^2\frac{u_xu_y}{\cosh^2ct}$,
\item[] $(x,y,t,u)\mapsto(\frac12e^{2x},y,t-x,u\,e^x)$ maps
$f=c^2\frac{(u_x+u_t+u)u_y}{\cosh^2c(x-t)}$ to
$f=c^2\frac{u_xu_y}{\cosh^2ct}$,
\item[]
$(x,y,t,u)\mapsto\bigl(\frac{x^{(c-1)/(c+1)}}{(c-1)},y,
\frac{t+\ln(x)}{(c+1)},u\,x^{-1/(c+1)}\bigr)$
maps $f=e^{ct}(xu_x-\frac{u_t+u}{c+1})$ to $f=e^{ct}u_x$.
 \end{itemize}

 The final list of integrable cases is summarised below.

 \begin{center}
\begin{tabular}{|l|l|l|}
 \hline
\# & Equation \eqref{DT} & Contact symmetry algebra $\g$\\
 \hline
1\vphantom{$\frac{\frac{A}A}{\frac{A}A}$} &
$f=c^2\frac{u_xu_y}{\cosh^2ct}$, $c>0$ & % parametrized by
2 functions of 1 variable \& 3 constants
 \\
 \hline
2\vphantom{$\frac{\frac{A}A}{\frac{A}A}$} &
$f=\frac{u_t^2-u^2}{(x-y)^2}$ &  % parametrized by
2 functions of 1 variable \& 3 constants \\
\hline
3\vphantom{$\frac{\frac{A}A}{\frac{A}A}$} &
$f=e^{ct}u_x$, $c\neq1$ &  % parametrized by
3 functions of 1 variable \& 2 constants\\
\hline
3$^+$\vphantom{$\frac{\frac{A}A}{\frac{A}A}$} &
$f=e^{t}u_x$ &  % parametrized by
3 functions of 1 variable \& 3 constants\\
\hline
4\vphantom{$\frac{\frac{A}A}{\frac{A}A}$} &
$f=e^{-t}(u_x+u_t+u)$ &  % parametrized by
3 functions of 1 variable  \& 2 constants\\
\hline
5\vphantom{$\frac{\frac{A}A}{\frac{A}A}$} &
$f=e^t(u_t-u)$ &  % parametrized by
3 functions of 1 variable  \& 2 constants\\
\hline
6\vphantom{$\frac{\frac{A}A}{\frac{A}A}$} &
$f= e^{ct}$, $c\ge0$ &  % possible split $2\neq c\geq0$ \& $c=2$
4 functions of 1 variable  \& 2 constants\\
\hline
\end{tabular}
\captionof{table}{Complete list of generalised DT integrable cases.}
 \end{center}

\medskip

Cases 1-6 are contact non-equivalent: this follows from the structure of  their contact symmetry algebras, where we use the notation of Section \ref{sec:wave}.

\medskip

(1) The generators are $Z_1(a)=a(x)\p_x$, $Z_2(b)=b(y)\p_y$,
$V_1=e^t\p_u$, $V_2=e^{-t}\p_u$, $V_3=u\p_u$.
Thus, $\mathfrak{s}_\infty=\op{Vect}(\R)\oplus\op{Vect}(\R)$,
$\mathfrak{s}_\diamond=\R\ltimes\R^2$ and
$[\mathfrak{s}_\infty,\mathfrak{s}_\diamond]=0$.

 \medskip

(2) Here $Z_1(a)=a(x)(\p_t-u\p_u)$, $Z_2(b)=b(y)(\p_t+u\p_u)$,
$V_1=\p_x+\p_y$, $V_2=x\p_x+y\p_y$, $V_3=x^2\p_x+y^2\p_y$.
Thus, $\mathfrak{s}_\infty=C^\infty(\R)\oplus C^\infty(\R)$,
$\mathfrak{s}_\diamond=\mathfrak{sl}(2)$ and
$[\mathfrak{s}_\infty,\mathfrak{s}_\diamond]=\mathfrak{s}_\infty$.

 \medskip

(3) Here $Z_1(a)=a(x)\p_x$, $Z_2(b)=b(y)\p_y-\frac{b'(y)}{c-1}(\p_t+u\p_u)$,
$Z_3(c)=c(y)e^{-t}\p_u$, $V_1=\p_t-cy\p_y$, $V_2=e^t\p_u$.
Thus, $\mathfrak{s}_\infty=\mathfrak{s}_\infty'\oplus\mathfrak{s}_\infty''$,
where $\mathfrak{s}_\infty'=\op{Vect}(\R)$ and
$\mathfrak{s}_\infty''=\op{Vect}(\R)\ltimes C^\infty(\R)$;
$\mathfrak{s}_\diamond=\R\ltimes\R=\op{sol}(2)$. In addition,
$[\mathfrak{s}_\infty,\mathfrak{s}_\diamond]=\mathfrak{s}_\infty''$
for $c\neq0$ and $[\mathfrak{s}_\infty,\mathfrak{s}_\diamond]=C^\infty(\R)
\subset\mathfrak{s}_\infty''$ for $c=0$.

\medskip

(3$^+$) Here $Z_1(a)=a(x)\p_x$, $Z_2(b)=b(y)(\p_t+u\p_u)$,
$Z_3(c)=c(y)e^{-t}\p_u$, $V_1=y\p_y-\p_t$, $V_2=\p_y$, $V_3=e^t\p_u$.
Thus, $\mathfrak{s}_\infty=\op{Vect}(\R)\oplus
(C^\infty(\R)\ltimes C^\infty(\R))$;
$\mathfrak{s}_\diamond=\R\ltimes\R^2$ and
$[\mathfrak{s}_\infty,\mathfrak{s}_\diamond]=
C^\infty(\R)\ltimes C^\infty(\R)$.

\medskip

(4) Here $Z_1(a)=a(x)(\p_x+\p_t-u\p_u)$,
$Z_2(b)=b(y)\p_y+\frac12b'(y)(\p_t+u\p_u)$,
$Z_3(c)=c(y)e^{-t}\p_u$, $V_1=\p_x$, $V_2=e^{t-2x}\p_u$.
Thus, $\mathfrak{s}_\infty=\mathfrak{s}_\infty'\oplus\mathfrak{s}_\infty''$,
where $\mathfrak{s}_\infty'=\op{Vect}(\R)$ and
$\mathfrak{s}_\infty''=\op{Vect}(\R)\ltimes C^\infty(\R)$;
$\mathfrak{s}_\diamond=\op{sol}(2)$ and
$[\mathfrak{s}_\infty,\mathfrak{s}_\diamond]=\mathfrak{s}_\infty'$.

\medskip

(5) Here $Z_1(a)=a(x)\p_x-\frac12a'(x)(\p_t-u\p_u)$,
$Z_2(b)=b(x)e^{t}\p_u$,
$Z_3(c)=c(y)(\p_t+u\p_u)$, $V_1=y\p_y-\p_t$, $V_2=\p_y$.
Thus, $\mathfrak{s}_\infty=\mathfrak{s}_\infty'\oplus\mathfrak{s}_\infty''$,
where $\mathfrak{s}_\infty'=\op{Vect}(\R)\ltimes C^\infty(\R)$,
$\mathfrak{s}_\infty''=C^\infty(\R)$;
$\mathfrak{s}_\diamond=\op{sol}(2)$ and
$[\mathfrak{s}_\infty,\mathfrak{s}_\diamond]=C^\infty(\R)'\oplus C^\infty(\R)''$.

\medskip

(6: $c\neq 2$) Here $Z_1(a)=a(x)\p_x-\frac{a'(x)}{c+2}(\p_t-u\p_u)$,
$Z_2(b)=b(x)e^{t}\p_u$,
$Z_3(c)=c(y)\p_y-\frac{c'(y)}{c-2}(\p_t+u\p_u)$, $Z_4(d)=d(y)e^{-t}\p_u$,
$V_1=x\p_x-y\p_y$, $V_2=\p_t+\frac{c}2u\p_u$.
Thus, $\mathfrak{s}_\infty=\mathfrak{s}_\infty'\oplus\mathfrak{s}_\infty''$,
where $\mathfrak{s}_\infty'=\op{Vect}(\R)\ltimes C^\infty(\R)=
\mathfrak{s}_\infty''$;
$\mathfrak{s}_\diamond=\R^2$ and
$[\mathfrak{s}_\infty,\mathfrak{s}_\diamond]=\mathfrak{s}_\infty$.

\medskip

(6: $c=2$) Here $Z_1(a)=a(x)\p_x-\frac{a'(x)}4(\p_t-u\p_u)$,
$Z_2(b)=b(x)e^{t}\p_u$,
$Z_3(c)=c(y)(\p_t+u\p_u)$, $Z_4(d)=d(y)e^{-t}\p_u$,
$V_1=y\p_y-x\p_x$, $V_2=\p_y$.
Thus, $\mathfrak{s}_\infty=\mathfrak{s}_\infty'\oplus\mathfrak{s}_\infty''$,
where $\mathfrak{s}_\infty'=\op{Vect}(\R)\ltimes C^\infty(\R)$,
$\mathfrak{s}_\infty''=C^\infty(\R)\ltimes C^\infty(\R)$;
$\mathfrak{s}_\diamond=\op{sol}(2)$ and
$[\mathfrak{s}_\infty,\mathfrak{s}_\diamond]=\mathfrak{s}_\infty$.

\medskip

Items 3 and 6 contain a parameter $c$ which is uniquely
characterised by the structure equations.

Item 1 also contains a parameter $c$, yet it does not enter the structure equations.
In this case non-equivalence does not follow from the symmetry analysis.
Instead we consider (point) transformations inducing an automorphism
of the symmetry algebra and preserving the orbit structure.
It is easy to see that such transformations,
leaving the class of Dunajski-Tod equations (\ref{DT}) form-invariant,
are only $(x,y,t,u)\mapsto(X(x),Y(y),t,ku)$ and so cannot change $c$.
Thus, the parameter $c$ is essential.

\medskip

\noindent{\bf Remark 1.}
A comparison between the two tables shows that a possible isomorphism
may exist for the following two cases:
 \begin{itemize}
\item
Table 1\,(1) to Table 2\,(1).
The symmetry algebras are abstractly isomorphic, yet the corresponding
two-dimensional subalgebras $[\mathfrak{s}_\diamond,\mathfrak{s}_\diamond]$ have orbits of dimensions 2 and 1 respectively, hence the items are not equivalent.
\item
Table 1(3) to Table 2(3) ($c\neq0$).
The symmetry algebras are abstractly isomorphic, yet the
corresponding infinite-dimensional subalgebras
$\mathfrak{s}'_\infty$ have orbits of dimensions 2 and 1 respectively,
hence the items are not equivalent.
 \end{itemize}
Thus, all integrable equations from Section \ref{sec:ex} (Tables 1-2 and BF equation) are pairwise
contact non-equivalent.

\bigskip

\noindent {\bf Remark 2.} For items 2, 3$^+$, 5, 6($c=2$) of Table 2,
the infinite part of the symmetry algebra is not perfect:
$[\mathfrak{s}_\infty,\mathfrak{s}_\infty]\subsetneqq\mathfrak{s}_\infty$.
Yet a closer analysis shows that the splitting and
the numerical characteristics of Table 2 are invariantly defined.

\bigskip

\noindent {\bf Remark 3.} The generalised Dunajski-Tod equation is quasi-linearisable: the contact transformation
 $$
\Phi(x,y,t,u,u_x,u_y,u_t)=
\left(x,y,\frac12\ln u_t,\frac{u-tu_t}{\sqrt{u_t}},
\frac{u_x}{\sqrt{u_t}},\frac{u_y}{\sqrt{u_t}},
\frac{-u-tu_t}{\sqrt{u_t}}\right)
 $$
maps equation \eqref{DT} to the quasilinear equation
 \begin{equation}\label{eqB}
u_xu_{yt}-u_tu_{xy}=h(x,y,t,u,u_x,u_y,u_t)\, u_{tt}
 \end{equation}
where $h=\tfrac14\Phi^*(f)$. Equation (\ref{eqB})  can be viewed as a
deformation of the Bogdanov equation \cite{Bogdanov}.
%Note that for $f=0$ we obtain the equation $\bigl(\frac{u_y}{u_t}\bigr){}_x=0$ which is easy to solve: $u=q(x,r(y,t))$, yet it is degenerate contrary to the cases $f\neq0$.
In the case $f=4e^{2\rho t}$ considered by Dunajski-Tod \cite{Dun6},
equation \eqref{eqB} becomes the integrable PDE studied in
\cite{Bogdanov}:
 $$
u_xu_{yt}-u_tu_{xy}=u_t^{\rho} u_{tt}.
 $$
The conformal structure for equation \eqref{eqB} is represented by the metric
 $$
g=4hdxdy+u_t^{-1}(u_xdx+u_tdt)^2,
 $$
and the corresponding Weyl covector is given by the
formula
 \begin{equation}\label{EWforB}
\omega=\Bigl(2\frac{u_x}{u_t}\mathcal{D}_t\ln h-\mathcal{D}_t\frac{u_x}{u_t}\Bigr)\,dx
+\mathcal{D}_t\ln u_y\,dy+2\mathcal{D}_t\ln h\,dt.
 \end{equation}
 Note that this Weyl covector satisfies formula \eqref{omega},
i.e.\ no `correction' is required, while the covector $\omega$
for generalised Dunajski-Tod equation does not satisfy \eqref{omega}, with
the `correction' being given by the second term containing $R$ in \eqref{EWforDT}.
This demonstrates contact non-invariance of $\omega$ given
by \eqref{omega}, while the covector $\omega=\Omega+\phi$ given by
Theorem \ref{t1} is genuinely contact invariant.

\medskip

The dispersionless Lax pairs for both generalised Dunajski-Tod \eqref{DT}
and generalised Bogdanov \eqref{eqB} equations can be obtained by the
recipe from the proof of Theorem \ref{t2}.
For the former, see Example 3 of \S\ref{sexs}.
This implies the Lax pair for the latter via the contact transformation $\Phi$.

\section {Reconstruction of  the Weyl covector}
\label{sec:omega}

We begin by describing the constraints for a PDE to be of Monge-Amp\`ere type.

\subsection{The Monge-Amp\`ere property}
\label{sec:MA}

Recall that equation (\ref{F}) is said to be of Monge-Amp\`ere type if its left-hand side can be  represented as a linear combination of minors (of all possible orders) of the Hessian matrix of the function $u$ (with coefficients depending on the 1-jet of $u$).
Let us represent equation (\ref{F}) in evolutionary form
\begin{equation}
u_{00}=f(x^0, x^1, x^2, u, u_0, u_1, u_2, u_{01}, u_{02}, u_{11}, u_{12}, u_{22}).
\label{m0}
\end{equation}
To calculate the Weyl covector, we will need explicit differential constraints for the right-hand side $f$ that are equivalent to the  Monge-Amp\`ere property. These have only been known  in low dimensions \cite{Boillat, Ruggeri, Colin, Gutt}. In  full generality, they were obtained recently in \cite{FKN}. In 3D, the Monge-Amp\`ere conditions consist of two groups of equations for $f$.
First of all, for every  $i\in \{1,2\}$ one has the relations
 \begin{equation}\label{MA1}
f_{u_{ii}}f_{u_{0i}u_{0i}}+f_{u_{ii}u_{ii}}=0, ~~~ f_{u_{0i}}f_{u_{0i}u_{0i}}+2f_{u_{0i}u_{ii}}=0.
 \end{equation}
Secondly, for every pair of distinct indices $i\ne j\in\{1,2\}$ one has the relations
 \begin{equation}\label{MA2}
 \begin{array}{c}
f_{u_{0j}}f_{u_{0i}u_{0i}}+2f_{u_{0i}}f_{u_{0i}u_{0j}}+2f_{u_{0i}u_{ij}}+2f_{u_{0j}u_{ii}}=0,
 \\[1em]
f_{u_{ij}}f_{u_{0i}u_{0i}}+2f_{u_{ii}}f_{u_{0i}u_{0j}}+2f_{u_{ii}u_{ij}}=0,
 \\[1em]
f_{u_{jj}}f_{u_{0i}u_{0i}}+f_{u_{ii}}f_{u_{0j}u_{0j}}+2f_{u_{ij}}f_{u_{0i}u_{0j}}+2f_{u_{ii}u_{jj}}+f_{u_{ij}u_{ij}}=0.
 \end{array}
 \end{equation}
Due to the contact invariance of the Monge-Amp\`ere class, the system of nine relations (\ref{MA1})--(\ref{MA2})  is invariant
under arbitrary contact transformations.

\subsection{Proof of Theorem \ref{t1}}\label{pthm1}

Let us consider a second-order PDE in evolutionary form (\ref{m0}).
Note that if a particular equation under study is not evolutionary,
 it can be brought to evolutionary form via
a suitable linear change of the independent variables.
It will be convenient to rewrite Einstein-Weyl conditions (\ref{EW}) in terms of the Levi-Civita connection of the conformal structure $g$ (choose any representative of the conformal class):
 \begin{equation}\label{EWLC}
r_{ij}+\frac{1}{2}(\nabla_i\omega_j+\nabla_j\omega_i)-\frac{1}{4}\omega_i\omega_j
=\Lambda g_{ij}
 \end{equation}
where $\nabla$ denotes covariant differentiation in the Levi-Civita connection of  $g$, and $r_{ij}$ is the corresponding Ricci tensor (which is automatically symmetric), see \cite{Jones}. Since  $g$ depends on the
2-jet of the function $u$, the Ricci tensor $r_{ij}$ depends on the
4-jet of $u$. This implies that components $\omega_k$ must depend on the 3-jet of $u$, furthermore, the dependence of $\omega_k$ on the
third-order derivatives of $u$ must be affine.
%Thus, the left-hand side of (\ref{EWLC}) will be linear in the fourth-order derivatives,  and quadratic in the third-order derivatives of $u$. Below we will look at the corresponding terms separately.
Analysis of the dependence of the left-hand side of (\ref{EWLC}) on the fourth-order derivatives of $u$ suggests a substitution
 \begin{equation}\label{Om}
\omega_k=\Omega_k+\phi_k
 \end{equation}
where $\Omega_k$ is given by  formula (\ref{omega}),
%$$ \Omega_k=2g_{kj}\mathcal{D}_{x_s}(g^{js})+\mathcal{D}_{x_k}(\ln\det g_{ij}),$$
and the `correction terms' $\phi_k$ are some functions to be determined (we will see that they can only depend on the 2-jet of $u$). Under this substitution equations (\ref{EWLC}) take the form
 \begin{equation}\label{EWLC1}
\begin{array}{c}
r_{ij}+\frac{1}{2}(\nabla_i\Omega_j+\nabla_j\Omega_i)+\frac{1}{2}(\nabla_i\phi_j+\nabla_j\phi_i)
-\frac{1}{4}\Omega_i\Omega_j-\frac{1}{4}(\Omega_i\phi_j+\Omega_j\phi_i)-\frac{1}{4}\phi_i\phi_j=\Lambda g_{ij}.
\end{array}
 \end{equation}
Let us denote by $S$ the system obtained from (\ref{EWLC1}) by eliminating  $\Lambda$ and restricting the resulting five equations to solutions of  PDE (\ref{m0}), that is, reducing the result modulo (\ref{m0}) and its differential prolongation.
Equations of system $S$ possess terms of several different types:
(a) linear in the fourth-order derivatives of $u$,
(b) quadratic in the third-order derivatives of $u$,
(c) linear in the third-order derivatives of $u$, and
(d) depending on the 2-jet of $u$ only.
We will discuss them case-by-case below.

\medskip

\noindent{\bf (a) Terms linear in the fourth-order derivatives of $u$.}
There are two sources of such terms: expressions
$r_{ij}+\frac{1}{2}(\nabla_i\Omega_j+\nabla_j\Omega_i)$ and $\frac{1}{2}(\nabla_i\phi_j+\nabla_j\phi_i)$. Direct calculation  shows that all terms with fourth-order derivatives of $u$ coming from the expressions $r_{ij}+\frac{1}{2}(\nabla_i\Omega_j+\nabla_j\Omega_i)$ cancel out. Thus, the only source of such terms are expressions $\frac{1}{2}(\nabla_i\phi_j+\nabla_j\phi_i)$, and this implies that  $\phi_k$  must be functions of the 2-jet of $u$ only: $\phi_k=\phi_k(x^0, x^1, x^2, u, u_0, u_1, u_2, u_{01}, u_{02}, u_{11}, u_{12}, u_{22})$, recall that $u_{00}$ can be eliminated via (\ref{m0}).
In other words, ansatz (\ref{Om}) captures the dependence of $\omega$ on the third-order derivatives of $u$.
For several classes of (translationally invariant) second-order PDEs the terms $\phi_k$ vanish identically, however, they are not zero in general.
Under conformal rescalings $g\to \lambda g$ both covectors $\omega$ and $\Omega$ transform as $\omega\to\omega+d\ln\lambda$, $\Omega\to\Omega+ d\ln\lambda$. Thus, the covector $\phi=\phi_kdx^k$ is invariant with
respect to conformal rescalings.
 % Beware this does not mean it is conformally invariant!

\medskip

\noindent{\bf (b) Terms quadratic in the third-order derivatives of $u$.}
Such terms come from the expressions $r_{ij}+\frac{1}{2}(\nabla_i\Omega_j+\nabla_j\Omega_i)-\frac{1}{4}\Omega_i\Omega_j$, and do not involve $\phi_k$. Equating to zero the corresponding coefficients we obtain all third-order partial derivatives of the function $f$ with respect to the variables $u_{01}, u_{02}, u_{11}, u_{12}, u_{22}$, which identically coincide with the integrability conditions for Hirota-type equations
\begin{equation}\label{Hir-evol}
u_{00}=f(u_{01}, u_{02}, u_{11}, u_{12}, u_{22})
\end{equation}
obtained in \cite{FHK}. This leads to a somewhat surprising conclusion: taking an integrable equation (\ref{m0}) and `freezing' the 1-jet of $u$ (that is, giving the variables $x^0, x^1, x^2, u, u_0, u_1, u_2$ arbitrary constant values), we obtain an integrable Hirota type equation. Note that the generic integrable Hirota type equation is a highly transcendental object: it coincides with the equation of the genus three hyperelliptic divisor \cite{CF}.

\medskip

\noindent{\bf (c) Terms linear in the third-order derivatives of $u$.}
These terms come from the expressions
$r_{ij}+\frac{1}{2}(\nabla_i\Omega_j+\nabla_j\Omega_i)+\frac{1}{2}(\nabla_i\phi_j+\nabla_j\phi_i)-\frac{1}{4}\Omega_i\Omega_j-\frac{1}{4}(\Omega_i\phi_j+\Omega_j\phi_i)$. Each of the five equations of system $S$ has seven terms linear in the third-order derivatives $u_{011},  u_{012},  u_{022},  u_{111},  u_{112},  u_{122}, u_{222}$, recall that we work modulo (\ref{m0}) and its differential prolongation.
Equating the corresponding coefficients to zero gives 35 relations involving $\phi_k$ and their first-order derivatives with respect to $u_{01}, u_{02}, u_{11}, u_{12}, u_{22}$. Eliminating the derivatives of $\phi_k$ we obtain a system of 20 equations which are linear inhomogeneous in $\phi_k$
 % (unfortunately, it is not possible to write them down explicitly in a journal paper due to their complexity).
(we do not write the equations explicitly due to their complexity).
It is exactly at this step that we can determine $\phi$ (and hence $\omega$) in terms of the function $f$. It should be stressed that the linear system
of 20 equations for $\phi_k$ is  nontrivial only if equation (\ref{m0})
is {\it not of Monge-Amp\`ere type}: in this case the linear system  can be represented in matrix form $A\phi=B$
where $\phi=(\phi_0,\phi_1,\phi_2)^T$, $B$ is a vector with 20
components and $A$ is a $20\times3$ matrix, whose coefficients depend
linearly on the left-hand sides of the Monge-Amp\`ere conditions (\ref{MA1})--(\ref{MA2}). For equations of non-Monge-Amp\`ere type
the unknowns $\phi_k$ can be reconstructed uniquely because $A$ necessarily
contains a nonzero $3\times 3$ minor. This is equivalent to the condition
$\op{rank}(A)=3$, note that we do not require $\op{rk}(A|B)=3$ as in
the Cramer rule. Indeed, the entire set of EW conditions
(more precisely, the differential closure of this system)
decomposes into constraints on $\phi_k$ and equations not containing $\phi_k$; the latter are integrability conditions for \eqref{m0}. Thus, part of
the constraints $A\phi=B$ % will contribute
contributes to the integrability conditions for the function $f$.

\medskip

\noindent{\bf (d) Terms depending on the 2-jet of $u$.}
For Monge-Amp\`ere equations, both the matrix $A$ and the vector $B$ of the linear system
$A\phi=B$ vanish identically.
% (their coefficients can be expressed in terms of the Monge-Amp\`ere conditions from Section \ref{sec:MA}).
In this case  further analysis is required.
Constraints of the second order in $u$ involve the derivatives of $\phi_k$; this overdetermined system for $\phi$ is not in involution.
Generically, the differential closure provides more PDEs that can ultimately
lead to algebraic formulae for $\phi$ via a finite jet of $u$.
Numerous examples show that this is indeed the case, and that
the Weyl covector $\omega$ can be reconstructed in terms of the equation
even for generic Monge-Amp\`ere equations.
However, explicit conditions and demonstration of this is
outside the scope of our paper.

\medskip

This finishes the proof of Theorem \ref{t1}. \qed

\bigskip

Calculations described above to reconstruct the Weyl covector $\omega$
are implemented in a Mathematica program which is attached to this submission.
%All the details of large computations can be found there.

\subsection{Examples of computations}
To illustrate the  general procedure, let us go through steps (a)-(c) for the two particular classes. % examples below.
 % As our goal is a formula for $\omega$, we will skip step (d).

\bigskip
\noindent{\bf Example 1.} Let us begin with equations
\begin{equation}\label{latinv}
u_{tt}=f(x, y, t, u, u_x, u_y, u_t, u_{xy}),
\end{equation}
the evolutionary form of lattice equations from Section \ref{sec:lat}. The characteristic conformal structure is (set $b=u_{xy}$):
$$
g=4dxdy-f_{{b}}dt^2.
$$
We will assume $f_{bb}\ne 0$, otherwise the equation is of Monge-Amp\`ere type.

\medskip

\noindent{Step (a):} calculation of $\Omega$ using formula (\ref{omega}) gives
$$
\Omega=\mathcal{D}_x(\ln f_{b}) dx+\mathcal{D}_y(\ln f_{b})dy-\mathcal{D}_t(\ln f_{b})dt,
$$
so that our ansatz for $\omega$ is
$$
\omega=\Omega+\phi_1dx+\phi_2dy+\phi_3 dt
$$
where $\phi_k$ are functions of the 2-jet of $u$.

\noindent{Step (b):} here we obtain only one non-trivial equation:
$$
f_{bbb}=2\frac{f_{bb}^2}{{f_b}}.
$$

\noindent{Step (c):} eliminating the derivatives of $\phi_k$ with respect to the variables $u_{xx}, u_{xy}, u_{xt}, u_{yy}, u_{yt}$, we obtain a linear system for $\phi_k$ (which vanishes identically if $f_{bb}=0$, that is, if the original equation is of Monge-Amp\`ere type). In the case $f_{bb}\ne 0$ this system gives an explicit formula for $\phi$:
$$
\phi_1=0, \quad \phi_2=0, \quad \phi_3=-\frac{2}{3}f_{u_t}+\frac{8}{3}\hat{\mathcal{D}}_t(\log f_b)-\frac{4}{3}\hat{\mathcal{D}}_t(\log f_{bb}),
$$
thus leading to an explicit formula for the Weyl covector $\omega$. Here  $\hat{\mathcal{D}}_t$ is the truncated total $t$-derivative (all differentiations are with respect to the 1-jet variables only):
$$
\begin{array}{c}
\hat{\mathcal{D}}_t=\partial_t+u_t\partial_u+u_{xt}\partial_{u_x}+u_{yt}\partial_{u_y}+u_{tt}\partial_{u_t}.
\end{array}
$$
We will not continue with  step (d): according to Section \ref{sec:lat}, it would lead to a conclusion that any equation (\ref{latinv}) with EW property is point-equivalent to the BF equation $u_{tt}=\ln u_{xy}$.

\bigskip
\noindent{\bf Example 2.} Let us consider equations of the form
\begin{equation}\label{xxyy}
u_{tt}=f(x, y, t, u, u_x, u_y, u_t, u_{xx}, u_{yy}).
\end{equation}
 The characteristic conformal structure is (set $a=u_{xx}, \ c=u_{yy}$):
$$
g={f_c}dx^2+{f_a}dy^2-f_af_cdt^2.
$$
We will assume that at least one of the second-order derivatives
$f_{aa}, f_{ac}, f_{cc}$ is nonzero,
otherwise the equation is of Monge-Amp\`ere type
(note also that $f_a\neq0,f_c\neq0$ if the equation is nondegenerate).

\medskip

\noindent{Step (a):} calculation of $\Omega$ using formula (\ref{omega}) gives
$$
\Omega=2\mathcal{D}_x\left(\ln {f_a}\right) dx+2\mathcal{D}_y\left(\ln {f_c}\right) dy,
$$
so that our ansatz for $\omega$ is
$$
\omega=\Omega+\phi_1dx+\phi_2dy+\phi_3dt
$$
where $\phi_k$ are functions of the 2-jet of $u$.

\medskip

\noindent{Step (b):} here we obtain a system of PDEs for $f$ in the arguments $a, c$:
\begin{eqnarray}
&&f_{aaa}=f_{aa}\left(\frac{f_{ac}}{f_c}+\frac{f_{aa}}{f_a}\right),\quad
f_{aac}=f_{aa}\left(\frac{f_{cc}}{f_c}+\frac{f_{ac}}{f_a}\right),\notag \\
&&f_{acc}=f_{cc}\left(\frac{f_{cc}}{f_c}+\frac{f_{ac}}{f_a}\right),\quad
f_{ccc}=f_{cc}\left(\frac{f_{cc}}{f_c}+\frac{f_{ac}}{f_a}\right),\quad
f_{aa}f_{cc}=(f_{ac})^2.\notag
\end{eqnarray}
These equations can be explicitly solved (see \cite{FHK}, Section 3.1).

\medskip

\noindent{Step (c):} eliminating the derivatives of $\phi_k$ with respect to the variables $u_{xx}, u_{xy}, u_{xt}, u_{yy}, u_{yt}$, we obtain a linear system for $\phi_k$. The first few equations of this system are as follows:
 $$
\begin{array}{c}
f_af_{cc}\phi_1=f_{ac}\hat{\mathcal{D}}_x(f_c)-f_{cc}\hat {\mathcal{D}}_x(f_a), \qquad
f_cf_{aa}\phi_2=f_{ac}\hat{\mathcal{D}}_y(f_a)-f_{aa}\hat{\mathcal{D}}_y(f_c),
 \\[1em]
3f_af_cf_{cc}\phi_2=4f_af_{cc}\hat{\mathcal{D}}_y(f_c)+4f_cf_{cc}\hat{\mathcal{D}}_y(f_a)-4f_af_{c}\hat{\mathcal{D}}_y(f_{cc})+2f_af_{cc}f_{u_y},
 \\[1em]
3f_af_cf_{cc}\phi_3=4f_af_{cc}\hat{\mathcal{D}}_t(f_c)+4f_cf_{cc}\hat{\mathcal{D}}_t(f_a)-4f_af_{c}\hat{\mathcal{D}}_t(f_{cc})-2f_af_{cc}f_{u_t},\
\text{etc;}
\end{array}
 $$
here $\hat{\mathcal{D}}_x, \hat{\mathcal{D}}_y, \hat{\mathcal{D}}_t$ are the truncated total derivatives.
% (all differentiations are with respects to the first-order jet variables only): $$ \begin{array}{c} \hat{\mathcal{D}}_x=\partial_x+u_x\partial_u+u_{xx}\partial_{u_x}+u_{xy}\partial_{u_y}+u_{xt}\partial_{u_t},\\ \hat{\mathcal{D}}_y=\partial_y+u_y\partial_u+u_{xy}\partial_{u_x}+u_{yy}\partial_{u_y}+u_{yt}\partial_{u_t},\\ \hat{\mathcal{D}}_t=\partial_t+u_t\partial_u+u_{xt}\partial_{u_x}+u_{yt}\partial_{u_y}+u_{tt}\partial_{u_t}. \end{array} $$
% Note that this  system vanishes identically if $f_{aa}=f_{ac}=f_{cc}=0$, that is, if equation (\ref{xxyy}) is of Monge-Amp\`ere type. On the contrary, if at least one of  $f_{aa}, f_{ac}, f_{cc}$ is nonzero, we can explicitly determine  $\phi$ (and hence $\omega$).
In the non-Monge-Amp\`ere case we can explicitly determine $\phi$ (and hence $\omega$).
For instance, if $f_{cc}\ne 0$ then the first and the last two of the above equations give explicit values for $\phi_1, \phi_2, \phi_3$.

We will not continue with  step (d), which would eventually lead to a conclusion that any equation (\ref{xxyy}) with EW property is contact-equivalent to an integrable  Hirota type  equation of the form $u_{tt}=f(u_{xx}, u_{yy})$; see \cite{FHK}, Section 3.1, for a list of such equations.

\section{Dispersionless Lax pairs}
\label{sec:Lax}

A background solution is  the manifold $M=\R^3(x^0,x^1,x^2)$
or a domain thereof, equipped with a function $u$ solving \eqref{F}.
We encode it into the symbol $M_u$, which can be viewed as
$\op{graph}(u)\subset M\times\R$,
as well as its lift into the jet-space inheriting the geometric structure.
Of the latter we emphasise the characteristic variety, which is a
projectivisation of the null cone of $[g]$ at every point.
This bundle is  four-dimensional,  called the
correspondence space $\hat M_u$.

Recall that a dispersionless Lax pair (dLp) can be identified with
a rank 2 distribution $\hat\Pi$ in $\hat M_u$. The distribution $\hat \Pi$  depends on a finite jet of the solution $u$,
and is Frobenius integrable modulo  equation (\ref{F}).
The natural projection $\pi:\hat{M}_u\to M_u$ has projective fiber
$\mathbb{P}^1$ with coordinate $\lambda$ called the spectral parameter;
it parametrises null 2-planes $\Pi$
%$\Pi\in C^\infty(\hat{M},\pi^*\op{Gr}_2(TM))$
of the conformal structure $[g]$ on $M_u$.

It was shown in \cite{CalKrug} that modulo  equation (\ref{F})
such Lax pair is unique,
coisotropic with respect to the characteristic
variety,
%(the field of null cones of $g$),
and  the lift $\Pi\dashrightarrow\hat{\Pi}$ has the projective property.
%In addition, the dLp can be assumed normal off-shell: $[\hat{\Pi},\hat{\Pi}]=\pi_*^{-1}(\Pi)$ for the general $u$ that is not necessarily a solution to (\ref{F}).
In Lemma 4 of \cite{CalKrug} it was proved that the Weyl covector $\omega$
uniquely
%encoding the Weyl connection $\mathbb{D}$)
determines the lift (see also Lemma 5 of \cite{CalKrug}), however, no explicit formula
for the lift was provided. This is what we do below in the proof of Theorem \ref{t2}.

\subsection{Proof of Theorem \ref{t2}}\label{pT2}

Let $X,Y$ be $\lambda$-dependent vector fields generating $\Pi$,
and let
 $$
\hat{X}=X+m\p_\lambda,\quad \hat{Y}=Y+n\p_\lambda
 $$
be their lifts to $\hat{\Pi}$. A section $\lambda=\lambda(\bx)$
is foliated by a one-parametric family of integral surfaces
of $\hat{\Pi}$ iff
 $$
\hat{X}(\lambda-\lambda(\bx))=m-X(\lambda(\bx))=0,\quad
\hat{Y}(\lambda-\lambda(\bx))=n-Y(\lambda(\bx))=0.
 $$
 This gives
 \begin{equation}\label{mn}
m=X(\lambda(\bx)),\quad n=Y(\lambda(\bx)),
 \end{equation}
and it remains to show that all first-order derivatives of $\lambda$ on the right-hand sides of (\ref{mn})  can be eliminated.
Let $\theta\in\Pi^\perp$
be a ($\lambda$-dependent) annihilator of the 2-plane congruence $\Pi$.
The condition that the Weyl connection $ {\mathbb{D}}$ preserves the field of null cones is
 \begin{equation}\label{dLpW}
{\mathbb{D}}_X\theta\wedge\theta=0,\quad  {\mathbb{D}}_Y\theta\wedge\theta=0,
 \end{equation}
where we substitute $\lambda=\lambda(\bx)$ prior to differentiation.
This condition gives precisely two linearly independent equations on the
1-jet of $\lambda(\bx)$, and these imply that all derivatives of $\lambda$ on the right-hand sides of (\ref{mn}) cancel out, leading to the required formulae for $m$ and $n$.

%\noindent {\bf Remark.} The system of two PDEs for hypersurfaces $\lambda=\lambda(\bx)$, obtained in the proof, is compatible and has solutions depending on one  function of one variable. It corresponds to twistor curves -- the images of hypersurfaces in the twistor space under the quotient by the leaves of the foliation of $\hat{M}_u$ by the dLp foliation on-shell, i.e.\ for any solution $u$ of (\ref{F}).

Given $g, \,\omega$ and following the above scheme, let us derive an explicit formula for dLp.
  %In full generality, it has never appeared in the literature.
First of all, we choose a (nonholonomic) null coframe
$\theta^0,\theta^1,\theta^2$ such that
 $$
g=4\theta^0\theta^2-(\theta^1)^2.
 $$
%This is always possible, but the coframe is nonholonomic in general.
Let  $V_0, V_1, V_2$ be the dual frame, and
let $c_{ij}^k$ be the structure functions defined by the expansions
 $$
[V_i,V_j]=c_{ij}^kV_k\quad\Leftrightarrow\quad
d\theta^k=-\sum_{i<j}c_{ij}^k\theta^i\wedge\theta^j.
 $$
The 2-plane congruence is $\Pi=\langle X=V_0+\lambda V_1,\
Y=V_1+\lambda V_2\rangle$ and
$\theta(\lambda)=\theta_2-\lambda\theta_1+\lambda^2\theta_0$.
Representing  the Weyl covector $\omega$ in the form $\omega=\omega_i\theta^i$ we compute the Weyl connection ${\mathbb{D}}$:
 \begin{align*}
{\mathbb{D}}\theta^0 =&
(c_{02}^2+4\omega_0)\theta^0\otimes\theta^0
+(\tfrac12c_{12}^2+2\omega_1-\tfrac12c_{01}^0-\tfrac14c_{02}^1)\theta^0\otimes\theta^1\\
&+(\tfrac12c_{01}^0+\tfrac12c_{12}^2+2\omega_1-\tfrac14c_{02}^1)\theta^1\otimes\theta^0\\
&+(\tfrac14\omega_2-\tfrac12c_{12}^1)\theta^1\otimes\theta^1
+c_{02}^0\theta^2\otimes\theta^0+c_{12}^0\theta^2\otimes\theta^1,
  \\[8pt]
{\mathbb{D}}\theta^1 =&
-2c_{01}^2\theta^0\otimes\theta^0+\tfrac12\omega_0\theta^0\otimes\theta^1
+(c_{12}^2-c_{01}^0+4\omega_1-\tfrac12c_{02}^1)\theta^0\otimes\theta^2\\
&+(c_{01}^1+\tfrac12\omega_0)\theta^1\otimes\theta^0
+\tfrac12\omega_1\theta^1\otimes\theta^1
+(\tfrac12\omega_2-c_{12}^1)\theta^1\otimes\theta^2\\
&+(c_{12}^2+4\omega_1-c_{01}^0+\tfrac12c_{02}^1)\theta^2\otimes\theta^0
+\tfrac12\omega_2\theta^2\otimes\theta^1+2c_{12}^0\theta^2\otimes\theta^2,
  \\[8pt]
{\mathbb{D}}\theta^2 =&
-c_{01}^2\theta^0\otimes\theta^1-c_{02}^2\theta^0\otimes\theta^2
+(\tfrac12c_{01}^1+\tfrac14\omega_0)\theta^1\otimes\theta^1\\
&+(\tfrac14c_{02}^1+2\omega_1-\tfrac12c_{12}^2-\tfrac12c_{01}^0)\theta^1\otimes\theta^2\\
&+(\tfrac14c_{02}^1+2\omega_1+\tfrac12c_{12}^2-\tfrac12c_{01}^0)\theta^2\otimes\theta^1
+(4\omega_2-c_{02}^0)\theta^2\otimes\theta^2.
 \end{align*}
Our convention is ${\mathbb{D}}\theta^k=-\Gamma^k_{ij}\theta^i\otimes\theta^j$ $\Leftrightarrow$ ${\mathbb{D}}_{V_i}\theta^k=-\Gamma^k_{ij}\theta^j$.
The torsion-free condition is equivalent to
$\op{alt}({\mathbb{D}}\theta^k)=\frac12d\theta^k$, and we also have
${\mathbb{D}} g=\omega\otimes g$.
Finally, using \eqref{dLpW} we compute the dLp  to be
 $$
\hat{\Pi}=\langle\hat{X}=V_0+\lambda V_1+m\p_\lambda,\ \
\hat{Y}=V_1+\lambda V_2+n\p_\lambda\rangle,
 $$
with
 \begin{align*}
m=&
(\tfrac12c_{12}^1-\tfrac14\omega_2)\lambda^3
+(\tfrac12c_{02}^1-c_{12}^2-\tfrac12\omega_1)\lambda^2
+(\tfrac12c_{01}^1-c_{02}^2-\tfrac14\omega_0)\lambda-c_{01}^2,\\[8pt]
n=&
-c_{12}^0\lambda^3
+(\tfrac12c_{12}^1-c_{02}^0+\tfrac14\omega_2)\lambda^2
+(\tfrac12c_{02}^1-c_{01}^0+\tfrac12\omega_1)\lambda
+(\tfrac12c_{01}^1+\tfrac14\omega_0).
 \end{align*}
Let us also note, following \cite{CalKrug}, \S4.3, that the lift of
$W=V_0+2\lambda V_1+\lambda^2V_2$ does not depend on the Weyl connection
and equals $\hat{W}=W+\sigma\p_\lambda$ with
 $$
\sigma=m+n\lambda=
-c_{12}^0\lambda^4 +(c_{12}^1-c_{02}^0)\lambda^3
-(c_{01}^0-c_{02}^1+c_{12}^2)\lambda^2
+(c_{01}^1-c_{02}^2)\lambda-c_{01}^2.
 $$
This is related to the fact that $W$ is null and is therefore independent
of the choice of $\omega$. The lift of other vectors from $\Pi$ does depend
on $\omega$.

As the  covector $\omega$ is algebraically determined by the equation, the Lax pair is also explicitly determined, thus finishing the proof of Theorem \ref{t2}. \qed

\subsection{Examples of computations}\label{sexs}

Below we discuss several examples illustrating the calculations described in the proof.

\bigskip

\noindent{\bf Example 1.} For the dispersionless Kadomtsev-Petviashvili (dKP) equation, $u_{xt}=(uu_x)_x+u_{yy}$,
the conformal structure is % represented by
 $$
g=4dxdt-dy^2+4udt^2,
 $$
and the  Weyl covector is $\omega=-4u_xdt$.
The corresponding Weyl connection $\mathbb{D}$ is given by the following nontrivial relations:
 \begin{gather*}
{\mathbb{D}}_{\p_x}\p_t={\mathbb{D}}_{\p_t}\p_x={\mathbb{D}}_{\p_y}\p_y=u_x\p_x,\ \
{\mathbb{D}}_{\p_y}\p_t={\mathbb{D}}_{\p_t}\p_y=u_y\p_t+2u_x\p_y,\\
{\mathbb{D}}_{\p_t}\p_t=(u_t-2uu_x)\p_x+2u_y\p_y+3u_x\p_t.
 \end{gather*}
We have $\theta(\lambda)=dx+\lambda dy+(\lambda^2+u)dt$ and
 $$
\Pi=\op{Ann}(\theta)=\langle
X=\p_y-\lambda\p_x,\ Y=\p_t-(\lambda^2+u)\p_x\rangle.
 $$
Condition \eqref{dLpW} gives
 $$
\lambda_t=\lambda^2\lambda_x+\lambda u_x+u \lambda_x+u_y,\ \
\lambda_y=\lambda \lambda_x+u_x,
 $$
leading to the familiar dLp for the dKP equation:
$\hat{X}=X+m\p_\lambda$, $\hat{Y}=Y+n\p_\lambda$ with
 $$
m=\lambda_y-\lambda\lambda_x=u_x,\quad
n=\lambda_t-(\lambda^2+u)\lambda_x =\lambda u_x+u_y.
 $$

\bigskip

\noindent {\bf Example 2.} For the dispersionless lattice equations
 (4) we have
 $$
\theta_0=f_{u_{tt}}dy,\ \theta_1=dt,\ \theta_2=dx,
 $$
so that $c_{01}^0=\mathcal{D}_t\ln f_{u_{tt}}$,
$c_{02}^0=\mathcal{D}_x\ln f_{u_{tt}}$
and $\omega_1=\omega(\p_t)$ are the only nonzero entries in
the above formulae, giving
 $$
m=-\tfrac12\omega_1\lambda^2,\quad
n=-\mathcal{D}_x\ln f_{u_{tt}}\lambda^2+(\tfrac12\omega_1-
\mathcal{D}_t\ln f_{u_{tt}})\lambda.
 $$
This reproduces the Lax pair from Section \ref{sec:lat}.

\bigskip

\noindent {\bf Example 3.} For the generalised Dunajski-Tod equations
(\ref{DT}) it is convenient to  change the representative of the conformal class as follows:
%(the formula below works for $f<0$ or in the complex case; in the  real case $f>0$ one can use the change $x\mapsto-x$ instead)
 $$
g=4dxdy-\alpha^2, \text{ where }
\alpha=\frac1{\sqrt{-f}}\Bigl(
(u_{xt}-u_x)dx+(u_{yt}+u_y)dy+(u_{tt}-u)dt\Bigr).
 $$
Then $\omega$ is changed to the new Weyl covector
 $$
\omega_{\text{new}}=\omega-d\ln|f|=\omega_0\theta^0+\omega_1\theta^1+\omega_2\theta^2
 $$
where $\theta^0=dx$, $\theta^1=\alpha$, $\theta^2=dy$ is a null coframe.
The dual frame is
 $$
V_0=\p_x-\frac{u_{xt}-u_x}{u_{tt}-u}\p_t,\quad
V_1=\frac{\sqrt{-f}}{u_{tt}-u}\p_t,\quad
V_2=\p_y-\frac{u_{yt}+u_y}{u_{tt}-u}\p_t.
 $$
The coefficients $\omega_i= \omega_{\text{new}}(V_i)$ are given by
$$
\begin{array}{c}
\omega_0= 2(1+R)\frac{u_{xt}-u_x}{u_{tt}-u}-\mathcal{D}_x\ln|f|,\quad \omega_1= \frac{2R\sqrt{-f}}{u_{tt}-u}, \quad
\omega_2= 2(1+R)\frac{u_{yt}+u_y}{u_{tt}-u}-\mathcal{D}_y\ln|f|.
\end{array}
 $$
The only nonzero structure functions are
 $$
c_{01}^1=d\alpha(V_1,V_0),\ c_{02}^1=d\alpha(V_2,V_0),\ c_{12}^1=d\alpha(V_2,V_1),
 $$
thus giving the Lax pair
$$
\hat{X}=V_0+\lambda V_1+m\p_\lambda,\quad
\hat{Y}=V_1+\lambda V_2+n\p_\lambda,
 $$
where
 \begin{align*}
m=&
(\tfrac12c_{12}^1-\tfrac14\omega_2)\lambda^3
+(\tfrac12c_{02}^1-\tfrac12\omega_1)\lambda^2
+(\tfrac12c_{01}^1-\tfrac14\omega_0)\lambda,\\[8pt]
n=&
(\tfrac12c_{12}^1+\tfrac14\omega_2)\lambda^2
+(\tfrac12c_{02}^1+\tfrac12\omega_1)\lambda
+(\tfrac12c_{01}^1+\tfrac14\omega_0).
 \end{align*}

\section{Rigidity conjecture}
\label{sec:rig}

Consider equation (\ref{F}) which satisfies EW property. As explained in Section \ref{sec:omega}, step (b), freezing  in (\ref{F}) the 1-jet of $u$ we obtain an integrable Hirota-type equation. It was demonstrated in \cite{FHK} that the parameter space of integrable Hirota-type equations is $21$-dimensional, supplied with a locally free action of the $21$-dimensional equivalence group ${\bf Sp}(6,\R)$.  This implies the existence of a universal Hirota master-equation generating an open ${\bf Sp}(6,\R)$-orbit in the $21$-dimensional parameter space.
It was shown in \cite{FKr} that integrability of  Hirota type equations is equivalent to the EW property, with the Weyl covector $\omega$ given by formula (\ref{omega}).
Finally, it was proved in \cite{CF} that the Hirota master-equation, which is a highly transcendental object, coincides with the equation of the genus three hyperelliptic divisor. We will say that PDE (\ref{F}) is {\it generic} if  a Hirota-type equation obtained by freezing a {\it generic}
1-jet of $u$ belongs to the open orbit.
%We would like to formulate the following
\bigskip

{\it \noindent {\bf Conjecture.}  A generic second-order PDE (\ref{F}) satisfying EW property is equivalent to the Hirota master-equation via a suitable contact transformation. In other words, for a generic PDE satisfying EW property, all dependence on the 1-jets is not essential,
and can be eliminated by a change of variables.
In fact, examples suggest that the `genericity' assumption can be weakened, leading to the following stronger conjecture: if a second-order PDE (which is not Monge-Amp\`ere) satisfies EW property, then it is equivalent to a Hirota type equation via a  contact transformation.}

\bigskip
An illustration of the rigidity phenomenon is given by the classification result of Section \ref{sec:lat}, where it was shown that any equation of type (\ref{lat}) satisfying EW property  is reducible to the  BF equation $u_{xy}=e^{u_{zz}}$.
For Monge-Amp\`ere equations the conjecture is certainly not true: there exist  contact non-equivalent examples  satisfying EW property (see Sections \ref{sec:wave} and \ref{sec:DT}, and Concluding Remarks for a discussion).

Below we prove two rigidity-type results, which motivate the above
conjecture and demonstrate a technique that may be utilised in its proof
in full generality. The main tool is the existence of an open orbit in the solution space of some differential equations with respect to their point symmetry groups.
 %We believe that the technique used in the proofs will also apply to the general rigidity conjecture (as it utilises the existence of an open orbit in the solution space of certain differential equations under the  group of their point symmetries).

\subsection{Rigidity result 1}

Let us consider Lagrangians of the form
\begin{equation}
\label{L1}
\int u_xu_y\varphi(u_t)\, dx dy dt.
\end{equation}
 It was shown in \cite{FO} that the requirement of integrability (EW property) of the corresponding second-order Euler-Lagrange equation implies that the function $\varphi(z)$ satisfies a fourth-order ODE
 \begin{equation}\label{L3}
\varphi''''(\varphi^2\varphi''-2\varphi \varphi'^2)-9\varphi'^2\varphi''^2+2\varphi \varphi'\varphi''\varphi'''+8\varphi'^3\varphi'''-\varphi^2\varphi'''^2=0,
 \end{equation}
whose general solution is a modular form of weight one and level three known as the Eisenstein series $E_{1, 3}(z)$.

 \begin{prop}\po \label{RR1}
Every Lagrangian of the form
 \begin{equation}\label{L2}
\int u_xu_yf(t, u, u_t)\, dx dy dt,
 \end{equation}
whose Euler-Lagrange equation satisfies EW property, is equivalent to its undeformed version (\ref{L1}) via a change of variables. In other words, Lagrangian (\ref{L1}) is rigid within the class (\ref{L2}).
 \end{prop}

\medskip

%\centerline{\bf Proof of triviality of deformation (\ref{L2}):}

\begin{proof}
The Euler-Lagrange equation for Lagrangian (\ref{L2}) is
\begin{equation}\label{Lagr}
u_xu_yf_{u_t,u_t}u_{tt}+2fu_{xy}+2f_{u_t}u_yu_{xt}+2f_{u_t}u_xu_{yt}+u_xu_y(f_u+f_{u_t,t}+u_tf_{u_t,u})=0.
\end{equation}
The corresponding characteristic conformal structure is
$$
g=-(f_{u_t}u_xdx+f_{u_t}u_ydy-fdt)^2+2u_xu_y(2f_{u_t}^2-ff_{u_t,u_t}).
$$
Looking for $\omega$ in the form (\ref{Om}) and substituting into the Einstein-Weyl conditions we obtain
$$
\phi_1=-u_x\frac{f_{u_t}}{f}Z, \quad \phi_2=-u_y\frac{f_{u_t}}{f}Z, \quad \phi_3=Z,
$$
where
$$
Z=\frac{6f_uf_{u_t}-2f_{u_t}f_{u_t,t}+2(f_t+f_uu_t)f_{u_t,u_t}-2(2f+f_{u_t}u_t)f_{u_t,u}}{ff_{u_t,u_t}-2f_{u_t}^2},
$$
(confirming that $\omega$ can be expressed in terms of the equation). Furthermore, we obtain four differential constraints for the function $f(t, u, u_t)$: one of them coincides with the ODE (\ref{L3}) in the variable $u_t=z$, while the other three are more complicated.  Utilising  ${\rm GL}(2)$-invariance of ODE (\ref{L3}) \cite{FO}, we look for a general solution in the form
$$
f(t, u, u_t)=\frac{q}{\gamma u_t+\delta}\ \varphi\left( \frac{\alpha u_t+\beta}{\gamma u_t+\delta}\right),
$$
where $\varphi(z)$ is a generic solution of (\ref{L3}), and $\alpha, \beta, \gamma, \delta, q$ should be considered as functions of the remaining arguments $t, u$. Under this ansatz, the ODE (\ref{L3}) (in the variable $u_t=z$) will be automatically satisfied. Direct analysis of the remaining constraints reveals that there exists a common factor $p$ such that $(p\alpha)_t=(p\beta)_u$ and $ (p\gamma)_t=(p\delta)_u$. Introducing the potentials one can set
$$
f(t, u, u_t)=\frac{q}{h_uu_t+h_t}\ \varphi\left( \frac{g_uu_t+g_t}{h_uu_t+h_t}\right),
$$
where  $q$ can be reconstructed uniquely up to a constant factor, leading to the following final answer:
\begin{equation}\label{fdef}
f(t, u, u_t)=\frac{(g_uh_t-h_ug_t)^2}{h_uu_t+h_t}\ \varphi\left( \frac{g_uu_t+g_t}{h_uu_t+h_t}\right);
\end{equation}
here $h(t, u)$ and $g(t, u)$ are two arbitrary functions. With any $f(t, u, u_t)$ given by (\ref{fdef}), equation (\ref{Lagr}) possesses EW property.
It is a common phenomenon that arbitrary functions occuring in the coefficients of integrable systems can be eliminated by a change of variables. This is exactly the case in our example: introducing the point transformation
$$
X=x, \quad Y=y, \quad T=h(t, u), \quad U=g(t, u),
$$
one can show that the densities
$$
U_XU_Y\varphi(U_T)\, dX\wedge dY\wedge dT  \quad {\rm and} \quad u_xu_yf(t, u, u_t)\, dx\wedge dy\wedge dt
$$
transform into each other, thus establishing triviality of deformation (\ref{L2}).
 \end{proof}

\subsection{Rigidity result 2}

 Equations of the form
\begin{equation}\label{Ch1}
u_{tt}=\frac{u_{xy}}{u_{xt}}+\frac{1}{6}\varphi(u_{xx})u_{xt}^{2}
\end{equation}
have appeared in the classification of integrable  hydrodynamic chains \cite{MaksEgor}, where it was shown that $\varphi$ must satisfy the Chazy equation \cite{Chazy} (set $u_{xx}=a$):
\begin{equation}\label{Ch}
\varphi_{aaa}+2\varphi \varphi _{aa}-3\varphi_a ^2=0.
\end{equation}

 \begin{prop}\po \label{RR2}
Every equation of the form
 \begin{equation}\label{Ch2}
u_{tt}=\frac{u_{xy}}{u_{xt}}+\frac{1}{6}f(x, u, u_x, u_{xx})u_{xt}^{2},
 \end{equation}
which satisfies  EW property,  is equivalent to its undeformed version (\ref{Ch1}) via a suitable contact transformation.
In other words, equation (\ref{Ch1}) is rigid within the class (\ref{Ch2}).
 \end{prop}

\medskip

% \centerline{\bf Proof of triviality of deformation (\ref{Ch2}):}

 \begin{proof}
The corresponding characteristic conformal structure is
$$
g=u_{xt}^3dxdy-\left(\frac{1}{6}f_a u_{xt}^6+\frac{1}{4}(u_{xy}-\frac{1}{3}f u_{xt}^3)^2\right)dy^2-\frac{1}{2}u_{xt}(u_{xy}-\frac{1}{3}fu_{xt}^3)dydt-\frac{1}{4}u_{xt}^2dt^2.
$$
Looking for $\omega$ in the form (\ref{Om}) and substituting into the Einstein-Weyl conditions we obtain
$$
\phi_1=\phi_3=0, \quad \phi_2=%-\frac{4}{3}u_{xt}^2u_tf_u
-\frac{2}{3}u_{xt}^3f_{u_x},
$$
(once again confirming that $\omega$ can be expressed in terms of the equation). Furthermore, we obtain the condition $f_u=0$, as well as four differential constraints for the function $f(x, u_x, u_{xx})$: one of them coincides with the Chazy equation (\ref{Ch}) in the variable $u_{xx}=a$, while the other three are more complicated.  Utilising  ${\rm SL}(2)$-invariance of the Chazy equation \cite{CO}, we look for a general solution in the form
$$
f(x, u_x, u_{xx})=\frac{1}{(\gamma u_{xx}+\delta)^2}\ \varphi\left( \frac{\alpha u_{xx}+\beta}{\gamma u_{xx}+\delta}\right)+\frac{6\gamma}{\gamma u_{xx}+\delta},\qquad \alpha \delta-\beta \gamma=1,
$$
where $\varphi(a)$ is a generic solution of (\ref{Ch}), and $\alpha, \beta, \gamma, \delta$ should be considered as functions of the remaining arguments $x, u_x$. Under this ansatz, the Chazy equation (\ref{Ch})  will be automatically satisfied. Direct analysis of the remaining constraints reveals that, analogously to the previous example, there exist potentials $g(x, u_x)$ and $h(x, u_x)$
such that $\alpha=g_{u_x}, \ \beta=g_x, \ \gamma=h_{u_x}, \ \delta=h_x$. This leads  to the following final answer:
\begin{equation}\label{fdef1}
f(x, u_x, u_{xx})=\frac{1}{(h_{u_x}u_{xx}+h_x)^2}\ \varphi\left( \frac{g_{u_x}u_{xx}+g_x}{h_{u_x}u_{xx}+h_x}\right)+\frac{6h_{u_x}}{h_{u_x}u_{xx}+h_x};
\end{equation}
here $g(x, u_x)$ and $h(x, u_x)$ are two functions which satisfy a single constraint $g_{u_x}h_x-h_{u_x}g_x=1$ (corollary of $\alpha \delta-\beta \gamma=1)$. With any $f(x, u_x, u_{xx})$ given by (\ref{fdef1}), equation (\ref{Ch2}) possesses EW property.
To eliminate arbitrary functions let us use the potential substitution $v=u_x$ which reduces equation (\ref{Ch1}) to  quasilinear form
\begin{equation}\label{Chq}
v_{tt}=\left(\frac{v_{y}}{v_{t}}+\frac{1}{6}f(x, v, v_x)v_{t}^{2}\right)_x.
\end{equation}
Equation (\ref{Chq}) can be equivalently represented as the condition of closedness of the 2-form
$$
v_t\ dx\wedge dy+\left(\frac{v_{y}}{v_{t}}+\frac{1}{6}f(x, v, v_x)v_{t}^{2}\right)dt\wedge dy.
$$
Introducing the point transformation
\begin{equation}\label{pt}
Y=y, \quad T=t, \quad X=h(x, v), \quad V=g(x, v),
\end{equation}
one can verify the identity
$$
\begin{array}{c}
V_T\ dX\wedge dY+\left(\frac{V_{Y}}{V_{T}}+\frac{1}{6}\varphi(V_X)V_{T}^{2}\right)dT\wedge dY
=v_t\ dx\wedge dy+\left(\frac{v_{y}}{v_{t}}+\frac{1}{6}f(x, v, v_x)v_{t}^{2}\right)dt\wedge dy,
\end{array}
$$
 thus demonstrating triviality of deformation (\ref{Ch2}) at quasilinear level (\ref{Chq}). The composition of  point transformation (\ref{pt}) with the potential substitution $v=u_x$ gives a contact transformation
 \begin{equation}\label{ct}
 \begin{array}{c}
 X=h(x, u_x), \ \ Y=y, \quad T=t, \ \ U=u+q(x, u_x), \ \
 U_X=g(x, u_x), \ \ U_Y=u_y, \quad U_T=u_t,
 \end{array}
 \end{equation}
where $q(x, u_x)$ is defined by the equations
$$
q_{u_x}=gh_{u_x}, \quad q_x=gh_x-u_x
$$
(which are compatible by virtue of the relation $g_{u_x}h_x-h_{u_x}g_x=1$). Contact transformation (\ref{ct}) takes equation (\ref{Ch2}) with $f(x, u_x, u_{xx})$ given by (\ref{fdef1}) to the undeformed equation
$$
U_{TT}=\frac{U_{XY}}{U_{XT}}+\frac{1}{6}\varphi(U_{XX})U_{XT}^{2},
$$
thus establishing the required triviality.
 \end{proof}

\subsection{ General rigidity conjecture}

The above arguments can be extended to the general case as follows.
Let $U$ be the Hessian matrix of a function $u$, and let
$F(U)=0$ be the Hirota master-equation. We will exploit two facts
(we change to complex coefficients for classification reasons):
 % changing to complex coefficients does not impact integrability
 \begin{itemize}
\item
Generic integrable Hirota-type equations belong to the same open
${\bf Sp}(6,\C)$-orbit, see \cite{FHK};
\item
Freezing 1-jet of $u$ in equation (\ref{F}) with EW property yields an integrable Hirota-type PDE, see \S\ref{pthm1}(b).
 \end{itemize}
Thus, a generic second-order PDE with EW property can be represented in the form
 \begin{equation}\label{Hir}
F\bigl((AU+B)(CU+D)^{-1}\bigr)=0
 \end{equation}
where $A, B, C, D$ are $3\times 3$ matrices depending on 1-jet variables $x^i,u,u_i$, such that the $6\times 6$ matrix
 $$
\left(\begin{array}{cc}
A&B\\
C&D\end{array}\right)
 $$
belongs to ${\bf Sp}(6,\C)$. Under the substitution (\ref{Hir}) part of the Einstein-Weyl conditions will be satisfied identically, leaving a (complicated!)  system of differential constraints for  $A,B,C,D$ in the
1-jet variables. It remains to prove that these constraints are equivalent to the existence of a contact  transformation taking PDE (\ref{Hir}) into  Hirota form $F(U)=0$. The complexity of the resulting differential constraints is a formidable obstacle in this programme.

\section{Concluding remarks}

{\bf 1.} In this paper we have studied second-order PDEs  in 3D whose characteristic conformal structure is Einstein-Weyl on every solution. A special subclass thereof are PDEs  whose characteristic conformal structure is {\it flat} on every solution (that is, has zero Cotton tensor). We
conjecture that any such PDE is contact equivalent to
$\triangle u=s$
where $\triangle$ denotes the Laplace operator of a constant-coefficient metric and $s$ is some function depending on the 1-jet of $u$. In other words, if the characteristic conformal structure is flat on every solution, the principal symbol can be reduced to  constant-coefficient form for all solutions simultaneously, via a suitable contact transformation. This result should be true in any dimension higher than  two (in two dimensions this is clearly false).

%\medskip \noindent {\bf 2.} We have shown that for a generic PDE  (\ref{F}) with Einstein-Weyl characteristic conformal structure there exists a  formula for covector $\omega$ in terms of the equation. In this connection, it would be interesting to construct explicit examples of non-flat conformal structures $[g]$  in 3D which can be turned into an Einstein-Weyl structure in several non-equivalent ways. This would imply that there exists no universal `formula' for $\omega$ in terms of $[g]$. We are not aware of such examples, see \cite{ET3} for a detailed discussion of Einstein-Weyl equations in 3D.

\medskip

\noindent {\bf 2.} We have demonstrated the existence of a  formula for the Weyl covector $\omega$ for PDEs (\ref{F})  that satisfy EW property and do not belong to the Monge-Amp\`ere class. This formula came from the terms  in the Einstein-Weyl equations that are linear in the third-order derivatives of $u$. We expect that analogous formula can be constructed for all
second order PDEs whose characteristic conformal structure $[g]$ is not
flat on generic solution (the required formula should follow from
the overdetermined system formed by the terms depending on 2-jet of $u$).
More generally, we conjecture that such a formula exists for
any PDE system whose characteristic variety is a nondegenerate quadric, yet
the corresponding conformal structure $[g]$ is not flat on  generic solution. This would include the Manakov-Santini system, thus agreeing with the results established in \cite{CalKrug}.

\medskip

\noindent {\bf 3.} It was observed in \cite{FKr} that second-order PDEs in 3D that are integrable by the method of hydrodynamic reductions \cite{FKh} must necessarily have EW property. Since  EW property (unlike hydrodynamic integrability) is manifestly contact-invariant,
it is tempting to adopt it as a contact-invariant approach to dispersionless integrability. This would  have a serious drawback: it is unknown at present how to solve such equations, indeed,  multiphase solutions coming from the method of hydrodynamic reductions may not be available.  On the other hand, we conjectured that non-Monge-Amp\`ere  second-order PDEs with EW property are in a sense `rigid': they can be reduced to  dispersionless Hirota form $F(u_{x^ix^j})=0$ via a suitable contact transformation. More generally, we expect that  Monge-Amp\`ere equations with Einstein-Weyl characteristic conformal structure (which  is not flat on every solution)
can be transformed, via a suitable B\"acklund transformation, into a translationally invariant form to which the method of hydrodynamic reductions would already apply.  Note that in the latter case contact transformations may not be sufficient.

As an illustrating example consider the following translationally non-invariant integrable deformation of the Veronese web equation,
\begin{equation}
(x^1-x^2)u_{x^3}u_{x^1x^2}+(x^2-x^3)u_{x^1}u_{x^2x^3}+(x^3-x^1)u_{x^2}u_{x^1x^3}=0;
\label{Ver}
\end{equation}
see \cite{KrugPan}, Theorem  8.1 for the corresponding Einstein-Weyl structure. Introducing the one-form
$$
\theta(\lambda)=(\lambda-x^2)(\lambda-x^3)u_{x^1}dx^1+(\lambda-x^1)(\lambda-x^3)u_{x^2}dx^2+(\lambda-x^1)(\lambda-x^2)u_{x^3}dx^3,
$$
(here $\lambda$ is a constant parameter), one can represent equation (\ref{Ver}) in compact form $d\theta(\lambda)\wedge \theta(\lambda)=0$.
% It gives rise to the Einstein-Weyl structure$$[g]=(x_2-x_3)^2u_1^2dx_1^2+(x_3-x_1)^2u_2^2dx_2^2+(x_1-x_2)^2u_3^2dx_3^2$$$$+2(x_3-x_1)(x_3-x_2)u_1u_2dx_1dx_2+2(x_2-x_1)(x_2-x_3)u_1u_3dx_1dx_3+2(x_1-x_2)(x_1-x_3)u_2u_3dx_2dx_3$$and covector$$\omega=$$ System (\ref{Ver}) possesses a Lax pair $[X, Y]=0$ where  $$ X=\partial_{x_1}-\frac{x_3-\lambda}{x_1-\lambda}\frac{u_1}{u_3}\partial_{x_3}, \qquad  Y=\partial_{x_2}-\frac{x_3-\lambda}{x_2-\lambda}\frac{u_2}{u_3}\partial_{x_3}. $$
Equation (\ref{Ver}) is not contact-equivalent to any translationally invariant equation. Indeed, if it was, it would possess a three-dimensional commutative subalgebra of contact symmetries (corresponding to translations in the independent variables). However, the contact symmetry algebra of equation (\ref{Ver}) is generated by vector fields
 $$
f(u){\partial}_u, \qquad
{\partial_{x^1}}+{\partial _{x^2}}+{\partial _{x^3}}, \qquad
x^1{\partial _{x^1}}+x^2{\partial _{x^2}}+x^3{\partial _{x^3}},
 $$
and one can easily see that this algebra does not contain any three-dimensional commutative subalgebra.
Nonetheless, equation (\ref{Ver}) is B\"acklund-related  to the translationally invariant Veronese web equation \cite{Z},
\begin{equation}
(a_1-a_2)u_{y^3}u_{y^1y^2}+(a_2-a_3)u_{y^1}u_{y^2y^3}+(a_3-a_1)u_{y^2}u_{y^1y^3}=0,
\label{Ver1}
\end{equation}
$a_i=\op{const}$,
which can be represented as $d\Theta(\lambda)\wedge \Theta(\lambda)=0$ where
$$
\Theta(\lambda)=(\lambda-a_2)(\lambda-a_3)u_{y^1}dy^1+(\lambda-a_1)(\lambda-a_3)u_{y^2}dy^2+(\lambda-a_1)(\lambda-a_2)u_{y^3}dy^3.
$$
 A B\"acklund transformation between equations (\ref{Ver}) and (\ref{Ver1})  can be represented in the form
\begin{equation}\label{theta}
\Theta(\lambda)=\theta(\lambda).
\end{equation}
This is a nonlocal change of the independent variables $x\to y$, note that the dependent variable $u$ remains unchanged. Setting in (\ref{theta}) successively $\lambda=a_i$ we obtain
\begin{equation}\label{recip}
\begin{array}{c}
u_{y^1}dy^1=\frac{\theta(a_1)}{(a_1-a_2)(a_1-a_3)}, \quad
u_{y^2}dy^2=\frac{\theta(a_2)}{(a_2-a_1)(a_2-a_3)},\quad
u_{y^3}dy^3=\frac{\theta(a_3)}{(a_3-a_1)(a_3-a_2)}.
\end{array}
\end{equation}
These relations specify the new independent variables $y^i$ uniquely modulo transformations $y^i\to \varphi^i(y^i)$, which are point symmetries of equation (\ref{Ver1}). B\"acklund transformation (\ref{recip}) can be viewed as a 3D version of reciprocal transformations that are well-studied in 2D.

\section*{Acknowledgements}

We thank D. Calderbank, B. Doubrov, M. Dunajski and M. Pavlov
for useful comments and clarifying discussions.
%The research of EVF was supported by the EPSRC grant EP/N031369/1.
%The research of VSN was supported by the EPSRC grant EP/V050451/1.
EVF and VSN  gratefully acknowledge financial support from a Russian Science Foundation grant (project 21-11-00006).
The work of BSK % and EVF
was partially supported by the project Pure Mathematics in Norway, funded by Trond Mohn Foundation and Troms\o\ Research Foundation.

\end{document}